\documentclass[superscriptaddress,aps,prX,twocolumn,showpacs,nofootinbib,longbibliography,notitlepage,floatfix]{revtex4-2}
\usepackage[utf8]{inputenc}
\usepackage{mathtools}
\usepackage{amsbsy}
\usepackage{physics}
\usepackage{amstext}
\usepackage{xcolor}
\usepackage{listings}
\usepackage{array}
\usepackage{svg}
\usepackage{siunitx}
\usepackage[dvipsnames]{xcolor}

\newcolumntype{C}[1]{p{#1}}

\usepackage[unicode=true,pdfusetitle,
 bookmarks=false,
 breaklinks=false,pdfborder={0 0 1},backref=false,colorlinks=false]
 {hyperref}

\usepackage{etoolbox}
\usepackage[english]{babel}
\usepackage{chngcntr}
\hypersetup{
 bookmarksnumbered=false,bookmarksopen=false}
\makeatletter

\@ifundefined{textcolor}{}
{
 \definecolor{BLACK}{gray}{0}
 \definecolor{WHITE}{gray}{1}
 \definecolor{RED}{rgb}{1,0,0}
 \definecolor{BLUE}{rgb}{0,0,1}
 \definecolor{CYAN}{cmyk}{1,0,0,0}
 \definecolor{MAGENTA}{cmyk}{0,1,0,0}
 \definecolor{YELLOW}{cmyk}{0,0,1,0}
 \definecolor{ORANGE}{rgb}{1,0.5,0}
}

\usepackage{amsmath}
\usepackage{graphicx}
\usepackage{amssymb}
\usepackage{txfonts,color}
\makeatother

\definecolor{codegreen}{rgb}{0,0.6,0}
\definecolor{codegray}{rgb}{0.5,0.5,0.5}
\definecolor{codepurple}{rgb}{0.58,0,0.82}
\definecolor{backcolour}{rgb}{0.95,0.95,0.92}

\lstdefinestyle{mystyle}{
  backgroundcolor=\color{backcolour}, commentstyle=\color{codegreen},
  keywordstyle=\color{magenta},
  numberstyle=\tiny\color{codegray},
  stringstyle=\color{codepurple},
  basicstyle=\ttfamily\footnotesize,
  breakatwhitespace=false,         
  breaklines=true,                 
  captionpos=b,                    
  keepspaces=true,                 
  numbers=left,                    
  numbersep=5pt,                  
  showspaces=false,                
  showstringspaces=false,
  showtabs=false,                  
  tabsize=2
}

\begin{document}


\title{Reducing Sensing Time through Offline Experimental Design for Nuclear Spin Detection}


\author{B. Varona-Uriarte}
\thanks{email: borjavarona201@gmail.com}
\affiliation{TECNALIA, Basque Research and Technology Alliance (BRTA),Parque Tecnológico de Bizkaia, C/ Geldo. Edificio 700, E-48160 Derio - Bizkaia (Spain)}
\affiliation{Department of Physical Chemistry, University of the Basque Country UPV/EHU, Apartado 644, 48080 Bilbao, Spain}
\affiliation{EHU Quantum Center, University of the Basque Country UPV/EHU, Leioa, Spain}

\author{F. Belliardo}
\affiliation{ 
    Institute of Photonics and Quantum Sciences, SUPA, School of Engineering and Physical Sciences, Heriot-Watt University, Edinburgh EH14 4AS, UK
}
\affiliation{ 
    Pritzker School of Molecular Engineering, University of Chicago, Chicago, Illinois 60637, USA
}

\author{M. H. Abobeih}
\affiliation{ 
    QuTech, Delft University of Technology, PO Box 5046, 2600 GA Delft, The Netherlands
}
\affiliation{ 
    Department of Physics, Harvard University, Cambridge, MA 02138, USA
}

\author{T. H. Taminiau}
\affiliation{ 
    QuTech, Delft University of Technology, PO Box 5046, 2600 GA Delft, The Netherlands
}
\affiliation{ 
    Kavli Institute of Nanoscience Delft, Delft University of Technology,
    PO Box 5046, 2600 GA Delft, The Netherlands
}

\author{C. Bonato}
\affiliation{ 
    Institute of Photonics and Quantum Sciences, SUPA, School of Engineering and Physical Sciences, Heriot-Watt University, Edinburgh EH14 4AS, UK
}

\author{E. Garrote}
\affiliation{TECNALIA, Basque Research and Technology Alliance (BRTA),Parque Tecnológico de Bizkaia, C/ Geldo. Edificio 700, E-48160 Derio - Bizkaia (Spain)}
\affiliation{Department of Automatic Control and Systems Engineering, University of the Basque Country UPV/EHU, 48013 Bilbao, Spain}

\author{J. Casanova}
\affiliation{Department of Physical Chemistry, University of the Basque Country UPV/EHU, Apartado 644, 48080 Bilbao, Spain}
\affiliation{EHU Quantum Center, University of the Basque Country UPV/EHU, Leioa, Spain}

\begin{abstract}
The characterization of nuclear spin environments in solid-state devices plays an important role in advancing quantum technologies, yet traditional methods often demand long measurement times.
To address this challenge, we integrate surrogate information gain (SIG) into our deep learning model based on the SALI architecture. By using SIG for data point selection, we achieve a significant reduction in experimental time while maintaining high precision in nuclear spin detection.
SIG is a figure of merit based on the expected variance of the signal, which is more straightforward to compute than the expected information gain (EIG) rooted in Bayesian estimation, and, crucially, it selects experiments that are more robust to experimental imperfections. We demonstrate our approach on a nitrogen-vacancy (NV) center in diamond coupled to $^{13}$C nuclei. In the high-field regime, our variance-based optimization is validated with experimental data, resulting in an 85$\%$ reduction in measurement time for a modest reduction in performance. In the low-field regime, we explore the model's performance on simulated data, predicting a 60$\%$ reduction in the total experimental time by improving the temporal resolution of the measurements and applying SIG. This demonstrates the potential of integrating deep learning with optimized signal selection to enhance the efficiency of quantum sensing and nuclear spin characterization, paving the way for scaling these techniques to larger nuclear spin systems.
\end{abstract}
\maketitle

\section{Introduction}

Quantum sensing experiments are fundamentally limited by the efficiency with which information can be extracted from a physical system. In many platforms, signal acquisition relies on single-photon detection and requires averaging over a large number of experimental repetitions to overcome shot noise, leading to long measurement times that restrict practical applicability~\cite{degen17, robledo11, zhang21}. Finding strategies to reduce acquisition time while preserving accuracy is therefore a central challenge for the broad deployment of quantum sensing technologies~\cite{awschalom18, gebhart23}.

A paradigmatic example of this challenge is the detection and characterization of individual nuclear spins~\cite{childress13, tisler13, nielsen17, vorobyov21}, a task of primary importance both for extending magnetic resonance imaging to the nanoscale~\cite{budakian24, du24, schwartz19} and for developing quantum technologies, where nuclear spins serve as long-lived auxiliary quantum registers for electron spin qubits~\cite{cramer16, abobeih22, reiner24}. Individual nuclear spins can be detected and coherently controlled via their hyperfine interaction with a single electron spin, for example associated to solid-state defects such as nitrogen-vacancy (NV) centers~\cite{bradley22, vanommen23}, silicon-vacancy centers~\cite{maity22}, tin-vacancy centers in diamond~\cite{beukers25}, and divacancies in silicon carbide~\cite{bourassa20}. 

Traditional approaches for nuclear spin cluster characterization rely on long pulse sequences applied to an electron spin to achieve high spectral resolution, resulting in acquisition times that restrict the broad applicability of these techniques~\cite{zhao12, kolkowitz12, taminiau12, fazhan14, muller14, liu16, zopes18, abobeih18, abobeih19, jung21}, particularly when one aims to screen a large number of systems with high throughput~\cite{galli25}. Efficient measurements that preserve accuracy are therefore essential for extending characterization techniques to larger and more complex quantum systems~\cite{bradley19, zahedian24}.

A promising avenue to reduce measurement time is to identify optimal data acquisition strategies that prioritize experimental settings expected to deliver the highest information gain~\cite{gebhart23}. These strategies can be non-adaptive, utilizing known prior information about the physics of the system, or can adaptively select the optimal data points based on the outcomes of previous measurements~\cite{joas21, arshad24, belliardo23, belliardo24}. 
In this work, we develop a non-adaptive strategy for the selection of measurement data points and benchmark it on an NV sensor.
Our approach is built on our recently developed deep learning framework for nuclear spin identification (SALI~\cite{varona24}), which serves as the underlying architecture for processing the experimental signals.

In contrast to Bayesian approaches~\cite{belliardo25}, pre-trained AI architectures such as SALI are tailored to a specific set of measurements and are therefore not straightforward to integrate with sequential adaptive experimental design. They are, however, advantageous due to their microsecond-scale run time. We show how \textit{offline} experimental design --with measurement settings optimized in advance rather than adapted during the experiment-- can still be highly beneficial, exploiting all available prior information on the target parameters to substantially reduce the resources needed for nuclear spin characterization.

Within the framework of Bayesian experimental design, the most well-motivated and widely used criterion is the expected information gain (EIG), defined as the mutual information between the unknown parameters and the measurement outcomes for a given experimental setting. Here, we introduce an alternative figure of merit, the surrogate (expected) information gain (SIG), which is computationally tractable and favors experimental settings that are more robust to measurement imperfections. We argue that EIG can identify experimental settings as informative even when, in practice, they yield little or no information due to unavoidable experimental errors, whereas SIG correctly identifies such settings as uninformative.

Our proposed estimation algorithm consists of three stages. In the first stage, we use the physical system model and a prior distribution of the target parameters, namely the hyperfine couplings between the nuclear spin and the NV center's electron spin, to determine the control settings most likely to provide the highest information gain, using the SIG as our figure of merit. In the second stage, we simulate measurements at these selected control settings and use the resulting dataset to train the SALI-architecture-based model, which, in contrast to conventional methods that rely on fitting resonances~\cite{shi14}, can operate on arbitrary input data. In the third stage, the SALI-based model is applied to real experimental data collected under the same conditions as the simulated training set.

Our work is organized as follows. In Sec.~\ref{sec:informationgain}, we present our offline experimental design methodology for reducing measurement time, introducing the concept of surrogate information gain as a computationally tractable, variance-based strategy rooted in general statistical principles. In Sec.~\ref{sec:derivation}, we analyze the relationship between EIG and SIG. In Sec.~\ref{experimentalvalidation}, we test the SALI-based model on experimental data to demonstrate its effectiveness in real scenarios. In Sec.~\ref{experimentaltimehighfield}, our methodology for measurement time reduction presented in Sec.~\ref{sec:informationgain} is validated on experimental data in the high-field regime, where the nuclear Larmor frequency $\omega_\mathrm{L}$ is much larger than the hyperfine couplings; we also propose a second time-saving strategy that reduces the number of repetitions used to compute each averaged data point, demonstrating an $85\%$ reduction in the total measurement time. In Sec.~\ref{experimentaltimelowfield}, we validate our methodology in the low-field regime through simulated data, predicting a 60\% reduction in measurement time.

\section{Experimental design and surrogate information gain}
\label{sec:informationgain}

We consider a quantum sensor based on a single electron spin ($S=1$), associated to an NV center in diamond, coupled to $n$ $^{13}\mathrm{C}$ nearby nuclear spins via the hyperfine coupling $\vec{A}_j=(A_j^{z},A_j^{\bot})$, where $j$ indexes the $n$ nuclear spins in the cluster. The NV electron spin, optically polarized, is prepared in an equal superposition state $\ket{+} \equiv \left(\ket{m_s=0} +\ket{m_s=1} \right)/\sqrt{2}$. We then apply a dynamical decoupling sequence $\left( \tau - \pi - \tau \right)$ repeated $\mathrm{N}$ times (CPMG sequence), where ``$\pi$'' represents a $\pi$-pulse on the electron spin, and $\tau$ the inter-pulse delay, followed by single-shot optical spin readout. By repeating this process $\mathrm{N}_{\mathrm{m}}$ times, one can experimentally measure the survival probability $P_+(\tau|\vec{A})$ of the initial state. We indicate the complementary probability of measuring the state $\ket{-} \equiv \left(\ket{m_s=0} - \ket{m_s=1} \right)/\sqrt{2}$ with $P_{-} (\tau|\vec{A})\equiv 1 - P_+(\tau|\vec{A})$. Together, these two probabilities define $P_\pm(\tau|\vec{A})$. The total measurement time required to collect sufficient data to estimate the experimental signal $P_+(\tau|\vec{A})$ can be approximated as $t = 2\mathrm{N} \cdot \mathrm{N}_{\mathrm{m}}\sum_j \tau_j$ for a sequence of $\mathrm{N}$ $\pi$-pulses and inter-pulse times $\tau_j$. Note that initialization and readout durations are neglected, as they are much shorter: sequences last milliseconds, while initialization and readout take only microseconds.

Experimental design consists of choosing the controllable parameters in an experiment to maximize the information we gain from performing it. The most common approach to experimental design in the context of Bayesian estimation is rooted in information theory and is based on the measurement of those data points that maximize the expected information gain (EIG)~\cite{rainforth24, lindley56}. The EIG is defined as the difference in entropy $S$ in the distribution of the unknown parameters (in our case, the vector of hyperfine parameters $\vec{A}\equiv\vec{A}_j$, containing the coupling coefficients for all spins) before and after incorporating the results of the measurements:
\begin{equation}
    \text{EIG}(\tau) \equiv \mathbb{E} \Big \lbrace S \left[ p(\vec{A}) \right] - S \left[ p(\vec{A} | \pm, \tau) \right] \Big \rbrace \;.
    \label{eig}
\end{equation}
The distribution $p(\vec{A})$ is the prior on the parameters, while $p(\vec{A} | \pm, \tau)$ is the posterior, i.e. the distribution of $\vec{A}$ conditioned to the observation of a certain outcome $\pm$ for the chosen inter-pulse delay $\tau$. The expectation value is taken over the possible outcomes $\pm$, computed according to our current knowledge of the system, i.e. the prior $p(\vec{A})$. The outcome probabilities entering the expectation value are given by
\begin{equation}
    P_\pm (\tau) := \int d \vec{A} \, p(\vec{A}) \, P_\pm (\tau|\vec{A}) \; . \label{eq:beronoulli_expectation_values}
\end{equation}
The expression in Eq.~(\ref{eig}) corresponds to the mutual information between the parameter vector $\vec{A}$ and the measurement outcome $\pm$, conditioned on the experimental configuration $\tau$. In principle, evaluating the expected information gain directly requires computing the posterior $p(\vec{A} | \pm, \tau)$, which is a computationally expensive task when the dimensionality of $\vec{A}$ is large. However, for a small number of discrete measurement outcomes, it is possible to introduce efficient Monte Carlo expressions that approximate the EIG without requiring evaluation of the posterior. More generally, modern variational techniques make it easier to amortize the cost of evaluating the information gain using neural networks~\cite{sarra23, rainforth24}. However, even with these precautions, computing EIG$(\tau)$ for a complex experiment such as the one discussed here can be challenging. We therefore explore alternative approaches that avoid Bayesian or variational methods and instead rely on physically motivated approximations. A further complication in our application is that the number of spins in the cluster is not known a priori, and therefore neither is the dimensionality of the vector $\vec{A}$ (see also~\cite{belliardo25, galli25}).

Under these assumptions, we propose to build an information gain figure of merit from quantities we can easily access, i.e. the model for the signal and the prior on $\vec{A}$, which we refer to as surrogate information gain (SIG):
\begin{equation}
    \text{SIG}(\tau) \equiv \mathbb{E} \lbrace \text{Var}_{\vec{A} \sim p(\vec{A})} \left[  P_{\pm}(\tau | \vec{A}) \right] \rbrace \; .
\label{eq:surrogate_IG}
\end{equation}
In the above expression, the expectation value over the outcomes $\pm$ is computed using the Bernoulli probability distribution in Eq.~\eqref{eq:beronoulli_expectation_values}, and the variance is computed over an ensemble of hyperfine parameters distributed like the prior. In other words, for a given measurement configuration 
$\tau$, we evaluate the variance of the probability of observing certain outcomes, and then compute its expectation value over the possible outcomes. Both the variance over $\vec{A}$ and the expectation value over the results of the measurements can be approximated with a Monte Carlo simulation. In the following, we simplify Eq.~\eqref{eq:surrogate_IG} by exploiting the fact that the measurement outcome is binary. We explicitly write the expectation value over the measurement outcome:
\begin{widetext}
\begin{equation}
    \text{SIG}(\tau) = P_+ (\tau) \text{Var}_{\vec{A} \sim p(\vec{A})} \left[  P_{+}(\tau | \vec{A}) \right]  + \left[ 1 - P_+ (\tau) \right] \text{Var}_{\vec{A} \sim p(\vec{A})} \left[  1 - P_{+}(\tau | \vec{A}) \right] = \text{Var}_{\vec{A} \sim p(\vec{A})} \left[  P_{+}(\tau | \vec{A}) \right] \; ,
\end{equation} 
\end{widetext}
from which we see that the expectation value simplifies and the SIG is just the variance of the signal we measure, since the variance of $1 - P_{+}(\tau | \vec{A})$ is the same as the variance of $P_{+}(\tau | \vec{A})$. We stress that, although the measurement outcome is binary in the present application, the definition of SIG extends naturally to measurements with an arbitrary number of outcomes through a suitable generalization of Eq.~\eqref{eq:beronoulli_expectation_values}. More generally, SIG can be applied to any estimation framework that includes a control parameter, a measurement outcome, and a prior distribution.

We now briefly comment on the interpretation of the SIG. Intuitively, experiments with the highest SIG are those where the signal exhibits the largest variance across the parameter values $\vec{A}$ that are consistent with our prior knowledge. In such cases, the measurement outcomes are most sensitive to differences in $\vec{A}$, making the experiment more informative. Conversely, if for a given inter-pulse delay $\tau$ the variance of the signal is small, then this control adds little value: all plausible values of $\vec{A}$ produce nearly identical signals, so the experiment cannot further refine our estimate. Finally, we observe that, due to the use of $P_+ (\tau|\vec{A})$ instead of $\log P_+ (\tau|\vec{A})$ and the non-linearity of the variance, the SIG is not an extensive quantity --that is, it does not increase proportionally as more measurement data are added.

\section{Relationship between EIG and SIG}
\label{sec:derivation}
In this section, we discuss more formally the relationship between EIG and SIG, and motivate the claim that SIG enables more robust selection of experiments. Assuming that the experiment has a finite number of possible outcomes, which makes it more tractable computationally, one can construct a Monte Carlo estimator for the EIG, which can be further improved through Rao-Blackwellization~\cite{rainforth24, vincent17, foster21}. For a two-outcome experiment, the Monte Carlo expression for the EIG becomes
\begin{equation}
    \text{EIG}(\tau) = h \left( P_{+}(\tau) \right) - \mathbb{E}_{\vec{A}\sim p(\vec{A})} \left[h( P_{+}(\tau | \vec{A}) ) \right] \;,
    \label{eq:rao_black_eig}
\end{equation}
where $h(x) \equiv - x \log_2 x - (1-x) \log_2 (1-x)$ is the binary entropy function, $P_{+}(\tau | \vec{A})$ is the outcome probability for a certain experiment, and $P_{+}(\tau)$ is the outcome probability integrated over the prior, defined in Eq.~\eqref{eq:beronoulli_expectation_values}. Intuitively, the expression in Eq.~\ref{eq:rao_black_eig} is the difference between the binary entropy of a Bernoulli variable with probability $P_{\pm}(\tau)$, and the average over the prior of the entropy of a Bernoulli variable with probability given by the survival probability $P_{+}(\tau | \vec{A})$. We now assume that the fluctuations of $P_{+}(\tau | \vec{A})$ over the prior distribution $p(\vec{A})$ are small, allowing us to perform a Taylor expansion of the binary entropy function as
\begin{equation}
    \begin{aligned}
    h(x) = h(x_0) &- \log_2 \left( \frac{x_0}{1-x_0} \right) (x-x_0) \\ &-\frac{1}{2 \ln 2} \frac{(x - x_0)^2}{x_0(1-x_0)} + o ( (x-x_0)^3) \; .
\end{aligned}
\end{equation}
This expression is valid for $x_0 \neq 0, 1$, where the binary entropy function is non-differentiable. Substituting it into Eq.~\eqref{eq:rao_black_eig}, we derive the following approximate equation for the EIG:
\begin{equation}
    \text{EIG}(\tau) \propto \frac{\text{Var}_{\vec{A} \sim p(\vec{A})} \left[  P_{+}(\tau | \vec{A}) \right]}{P_{+}(\tau) \left[ 1 - P_{+}(\tau) \right]} \; .
    \label{eq:normalized_sig}
\end{equation}
The right-hand side of Eq.~\eqref{eq:normalized_sig} is bounded by one because of the relation between the variance and the mean of a random variable in $[0, 1]$, like $P_{+}(\tau | \vec{A})$. This expression can be interpreted as a signal-to-noise ratio, where \textit{signal} refers to the fluctuation of the signal over the prior, and \textit{noise} refers to the shot noise associated to the measurement. In this sense, Eq.~\eqref{eq:normalized_sig} states that the information gain for a given measurement configuration is determined by the variance of the signal, rescaled to the shot noise level. This interpretation suggests that a large EIG can be achieved in a regime close to the extrema of the probability interval, i.e. $P_{\pm}(\tau) \simeq 0, 1$, where the shot noise becomes very small, even if the variance of the signal is also small. However, such a regime is generally unrealistic, since both the signal and the shot noise are likely to be dominated by background additive noise from the experimental apparatus. The restriction to discrete measurement outcomes, which allows the use of an efficient Rao-Blackwell estimator, also makes it difficult to introduce a more sophisticated noise model based on the addition of Gaussian fluctuations to the measurement outcomes. Intuitively, one would incorporate the strength of the additive Gaussian noise $\sigma$ as follows:
\begin{equation}
    \text{EIG}(\tau) \propto \frac{\text{Var}_{\vec{A} \sim p(\vec{A})} \left[  P_{+}(\tau | \vec{A}) \right]}{P_{+}(\tau) \left[ 1 - P_{+}(\tau) \right] + \sigma^2} \; .
\end{equation}
In the limit in which the noise floor is much larger than the shot noise, we obtain
\begin{equation}
    \text{EIG}(\tau) \propto \frac{1}{\sigma^2} \text{Var}_{\vec{A} \sim p(\vec{A})} \left[  P_{+}(\tau | \vec{A}) \right] \; ,
\end{equation}
which is proportional to the SIG for the binomial outcome. This derivation motivates the definition in Eq.~\eqref{eq:surrogate_IG} and shows that SIG is expected to be more robust to noise than EIG in its simplest form for finite outcome systems. Although this derivation is carried out for the binomial outcome case, it can be extended in a similar way to multiple outcomes, providing justification for the general definition of SIG.

\section{Benchmarking nuclear spin estimation on the full experimental dataset} \label{experimentalvalidation}

In this section, we first benchmark the performance of the SALI-based model on the full experimental dataset~\cite{abobeih18, jung21} in the high-field regime (specifically at $B_z = 404\ \mathrm{G}$). The high-field regime, where the Larmor frequency of the nuclear spins is larger than their hyperfine couplings to the electron spin, is the operating condition of most experiments. Following this validation, in Secs.~\ref{experimentaltimehighfield} and~\ref{experimentaltimelowfield}, we demonstrate our approach for measurement time reduction.

\subsection{Model training} \label{modelinginputsignals}

To train the model, we generate a {\it synthetic} dataset by numerically simulating distinct $P_+ \equiv P_+(\tau, \vec{A})$ signals (see the explicit expression for $P_+$ in Sec.~\ref{equationderivation} of the Supplemental Material (SM)~\cite{supplementalmaterial}). All models in this work have been trained on a dataset of 5 million synthetic samples. These samples were generated based on a prior distribution of the hyperfine coupling constants, where the number of nuclear spins $n$ surrounding the NV detector was uniformly sampled between 1 and 50, and each spin was assigned coupling components $A^z$ and $A^{\bot}$ drawn from uniform ranges $A^z\in~[-50, \ 50] \ \mathrm{kHz}$ and $A^{\bot}\in~[2, \ 80] \ \mathrm{kHz}$, respectively.

To account for the effect on the input signals ($P_+$) of a weakly coupled spin-bath, comprising thousands of spins with couplings $A^{\bot}\in~[0, \ 2] \ \mathrm{kHz}$, we simulated a set of random crystal configurations and computed their respective spin-bath contributions. For each generated sample, one crystal configuration was randomly selected, and its spin-bath contribution was added to the $P_+$ signal. For details on the training process, refer to Sec.~\ref{straining} of the SM~\cite{supplementalmaterial}.

To accommodate the model to the available experimental data, we use as input two $P_+$ signals corresponding to pulse sequences with $\mathrm{N}=32$ $\pi$-pulses and inter-pulse spacing $\tau_{32}$ in the interval $[6,\ 50] \ \mathrm{\mu s}$, and $\mathrm{N}=256$ $\pi$-pulses over $\tau_{256}\in[10,\ 40] \ \mathrm{\mu s}$, with temporal resolutions of $\Delta \tau_{32} = \Delta \tau_{256} = 4\ \mathrm{ns}$. In addition, NV decoherence effects are modeled as $P_+=(1 + \mathrm{M} \cdot e^{-\tau / \mathrm{T}}) / 2$, where $\mathrm{M}$ represents the total contribution from all NV-nuclei couplings~\cite{jung21}. The coherence time $\mathrm{T}$ is computed as $\mathrm{T}^{\mathrm{N}}=\mathrm{T}^{\mathrm{N}=4}\cdot(\mathrm{N}/4)^{\eta}$, with $\mathrm{T}^{\mathrm{N}=4}=3\ \mathrm{ms}$ and $\eta=0.8$. This equation is fitted to experimental data in Ref.~\cite{abobeih18}, characterizing the dependence of coherence time on the number of pulses $\mathrm{N}$. Furthermore, shot noise is simulated by averaging each data point in $P_+$ over $\mathrm{N}_{\mathrm{m}}=250$ repetitions of the dynamical decoupling sequences, leading to a cumulative measurement duration of roughly 8 hours. The intervals for $\tau$, temporal resolutions $\Delta \tau$, and number of measurements per data point $\mathrm{N}_{\mathrm{m}}$ reported in this section are taken as the default hyperparameters of the model. Variations of these parameters are explored throughout the paper, as specified in each section.

The ground-truth outputs of the model are 2-dimensional images of size $204 \times 160$, where nuclear spins are represented by Gaussian peaks with unit variance, each peak’s position corresponding to the spin’s coupling constants: $A^z$ on the y-axis and $A^{\bot}$ on the x-axis.

\begin{figure*}[t]
\includegraphics[width=1\linewidth]{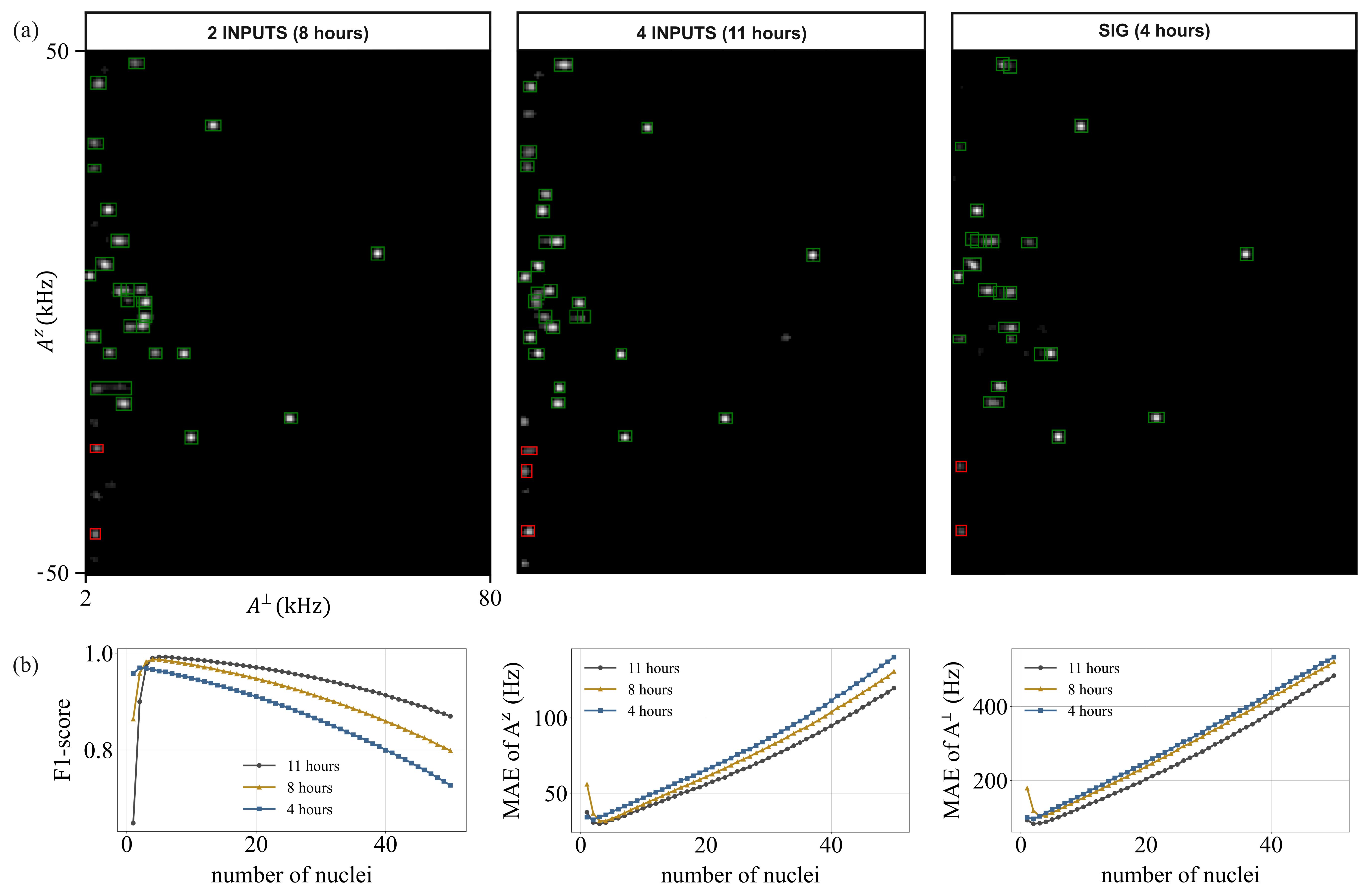}
\caption{(a) Outputs of the SALI-based model for three cases using: 2 input signals ($\mathrm{N}=32,\ 256$), 4 input signals ($\mathrm{N}=32,\ 96,\ 128,\ 256$), and the 2-input case with our method to reduce measurement time described in Sec.~\ref{varianceapproachhighfield}. Measurement times are 8 hours, 11 hours, and 4 hours, respectively. The y-axis represents the parallel component of the hyperfine coupling $A^z$, while the x-axis shows the perpendicular component $A^{\perp}$. In the first case, 30 nuclear spins are detected, of which 27 match the predictions from Ref.~\cite{jung21} (green boxes). In the second case, 29 spins are detected with 27 matches, and in the third, 27 are detected with 25 matches. Red boxes indicate spins not matching the reference predictions. (b) Performance metrics for the three models calculated across the test set, including F1-score and Mean Absolute Error (MAE) for both coupling components. As expected, reducing input signal information slightly degrades performance, but the impact remains minimal. Numerical values corresponding to these plots are provided in Tables~\ref{f1scorehighfield}, \ref{maeAhighfield}, and \ref{maeBhighfield} of Sec.~\ref{numericalresults} in the SM~\cite{supplementalmaterial}.}
\label{outputimages}
\end{figure*}

\subsection{Testing the model with experimental data} \label{testingmodel}
We deploy the trained model on the experimental data, obtaining the image on the left of Fig.~\ref{outputimages}~(a). A total of 29 nuclear spins are identified, 27 of which align with estimations using other techniques~\cite{jung21}. These 27 spins are shown in green boxes in the same image.
The remaining two spins, highlighted in red boxes, may correspond to a strongly coupled nuclear spin predicted in Ref.~\cite{jung21}, which lies outside our sampling range (beyond the image boundaries).
Details of the estimated nuclear spins are reported in Tables~\ref{predictions} and~\ref{predictionsnomatch} of Sec.~\ref{numericalpredictions} in the SM~\cite{supplementalmaterial}.

Leveraging the flexibility of the SALI architecture in handling a varying number of inputs (i.e., a different number of $P_+$ input signals), we evaluate its performance when dealing with four inputs. These include the previously used $P_+$ data strings, corresponding to sequences with $\mathrm{N}=32$ and $\mathrm{N}=256$ $\pi$-pulses, along with two additional experimental signals that emerge after driving the system with CPMG sequences containing $\mathrm{N}=96$ and $\mathrm{N}=128$ $\pi$-pulses, where $\tau_{96} = \tau_{128} \in [10,\ 30]\ \mathrm{\mu s}$ and $\Delta \tau_{96} = \Delta \tau_{128} = 4\ \mathrm{ns}$. Note that the model is retrained to accommodate four inputs. 
The resulting output is the middle image of Fig.~\ref{outputimages}~(a), and reveals 30 nuclear spins. Of these, 27 match previously detected spins (highlighted in green boxes), while the effect of the strongly coupled nuclear spin is also evident (nuclear spins highlighted in red boxes).

The performance of the models, shown in Fig.~\ref{outputimages}~(b), is evaluated using two key metrics. The first one is F1-score, which quantifies the accuracy of the number of detected nuclear spins. This is defined as $2(\mathrm{P} \cdot \mathrm{R})/(\mathrm{P} + \mathrm{R})$, where $\mathrm{P}$ is precision and $\mathrm{R}$ is recall. The second is the Mean Absolute Error (MAE) of the predicted coupling constants $A^z$ and $A^{\bot}$ for the detected nuclear spins, which assesses the accuracy of parameter estimation. The definitions of the metrics are provided in Sec.~\ref{metricsdescription} of the SM~\cite{supplementalmaterial}.

\section{Measurement time reduction: high field} \label{experimentaltimehighfield}

In this section, we benchmark the performance of our procedure for identifying the most informative data points in experimental signals, which are then processed using the SALI-based model. Practically, this allows experimentalists to measure only a subset of data points (those selected by our method), significantly reducing the total measurement time.

\begin{figure*}
\includegraphics[width=1 \linewidth]{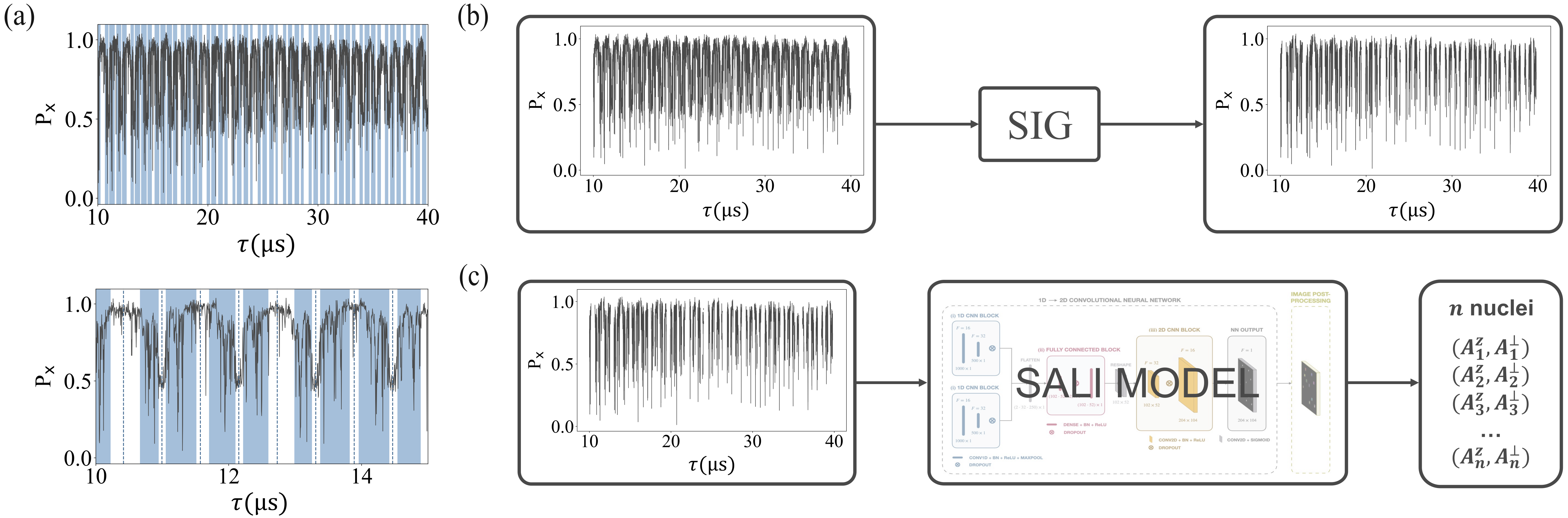}
\caption{(a) Top: Experimental signal corresponding to a CPMG pulse sequence with $\mathrm{N} = 256$ $\pi$-pulses. The blue fringes indicate the data points selected by the SIG approach, while the blank regions correspond to discarded points. Bottom: Zoomed-in view of the same signal in the region $\tau \in [10,\ 15]\ \mathrm{\mu s}$. The blue vertical dashed lines mark the inverse of the Larmor frequencies of the nuclear spins. The resonances around these frequencies correspond to the weakly coupled spin-bath ($A^{\bot} \in [0,\ 2]\ \mathrm{kHz}$), which lies outside our sampling range and therefore does not provide relevant information in our case.
(b) Application of the SIG approach: a subset of data points is selected from the signal.
(c) Operational model: The trained model, using only the selected data points, is applied to experimental signals. In this example, only the experimental signal corresponding to $\mathrm{N} = 256$ $\pi$-pulses is shown. The same procedure is also applied to the signal with $\mathrm{N} = 32$ $\pi$-pulses.}
\label{figure_variance}
\end{figure*}

\subsection{SIG approach: high field} \label{varianceapproachhighfield}

We consider here the high-field regime, where the nuclear spin Larmor frequency is larger than the hyperfine couplings to the central electron spin. In this regime, the signal consists of resonance dips whose positions are determined by the parallel component of the hyperfine coupling $A^z$, while their depths are related to the perpendicular component $A^{\bot}$. When multiple nuclear spins are coupled to the NV center, the resonance peaks can overlap in the experimental signal, so that different spins may contribute to a broader dip at the same value of $\tau$. These overlaps make interpretation more challenging, but they also highlight regions of the signal that contain richer information.

To identify these highly informative regions, we apply the SIG approach presented in Sec.~\ref{sec:informationgain}. Specifically, we compute the variance of the $P_+$ signal at each data point (i.e., at each $\tau$) across a batch of thousands of synthetic samples drawn from the same prior distribution described in Sec.~\ref{modelinginputsignals}. We then select the $\mathrm{N}_{\mathrm{p}}$ data points with the highest variance, prioritizing those that differ the most between samples. This ensures that the AI model relies on sample-specific data points, rather than common ones that do not contribute to distinguishing between different samples.

In Fig.~\ref{figure_variance}~(a), the top plot displays the experimental signal acquired using a CPMG sequence with $\mathrm{N} = 256$ $\pi$-pulses. The vertical blue fringes indicate the data points selected by the SIG approach. The bottom plot provides a zoomed-in view of the signal in the region $\tau \in [10,\ 15]\ \mathrm{\mu s}$, where vertical dashed lines mark multiples of $\frac{\pi}{2\omega_L}$, with $\omega_L$ being the Larmor frequency of the nuclear spins. Around odd multiples of $\frac{\pi}{2\omega_L}$, one can observe resonances from the weakly coupled spin-bath ($A^{\bot} \in [0,\ 2]\ \mathrm{kHz}$), which lies outside our target sampling range used in the computation of the SIG. Consequently, the data points near the Larmor frequencies do not contribute to our characterization task. This highlights that the selected data points effectively emphasize the most informative regions for identifying nuclear spins within the desired coupling ranges, $A^z \in [-50,\ 50]\ \mathrm{kHz}$ and $A^{\bot} \in [2,\ 80]\ \mathrm{kHz}$, as described earlier.
Fig.~\ref{figure_variance}~(b) shows the result of applying this selection method to the full experimental signal. The same approach is applied independently to the signal corresponding to $\mathrm{N} = 32$ $\pi$-pulses. In Fig.~\ref{figure_variance}~(c), the model is trained using a synthetic dataset generated by simulating $P_+$ signals only at the data points selected by the SIG approach. The resulting model --referred to as the operational model-- is then deployed to characterize nuclear spin systems from experimental signals.

\begin{figure*}[t]
\includegraphics[width=1 \linewidth]{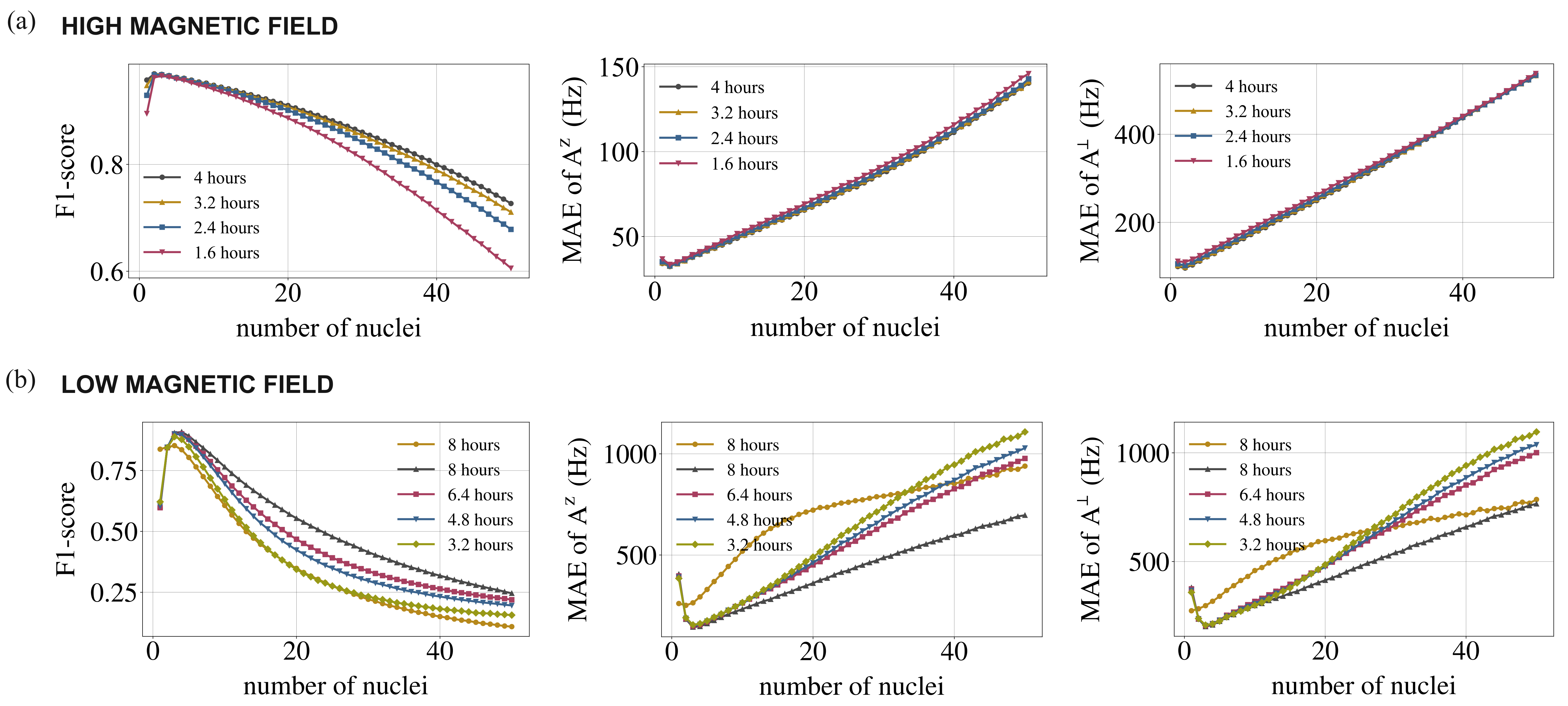}
\caption{Performance metrics demonstrating model robustness against shot noise in different magnetic field regimes. (a) High-field regime ($B_z = 404\ \mathrm{G}$): The 4-hour case corresponds to the model incorporating the SIG approach, with $\mathrm{N}_{\mathrm{m}}=250$ measurements per data point. The yellow, blue, and pink lines represent the same model tested with simulated measurements of $\mathrm{N}_{\mathrm{m}}=200$ (3.2 hours), $\mathrm{N}_{\mathrm{m}}=150$ (2.4 hours), and $\mathrm{N}_{\mathrm{m}}=100$ (1.6 hours), respectively. The 1.6-hour case achieves an $85\ \%$ reduction in measurement time compared to the 11-hour model trained with 4 input signals. The first three models exhibit nearly identical performance, whereas the last model shows slight degradation. Numerical values corresponding to these plots are provided in Tables~\ref{f1scorehighfield}, \ref{maeAhighfield}, and \ref{maeBhighfield} of Sec.~\ref{numericalresults} in the SM~\cite{supplementalmaterial}. (b)~Low-field regime ($B_z = 40.4\ \mathrm{G}$): The yellow line represents the reference model trained using the same 2 input signals as in the high-field case (see Sec.~\ref{modelinginputsignals}). The black line corresponds to the model trained on the data derived from the SIG approach. Although both cases have the same measurement time (8 hours), applying the SIG approach significantly improves model performance. The pink, blue, and green lines show performance under increasing shot noise, with $\mathrm{N}_{\mathrm{m}}=200$ (6.4 hours), $\mathrm{N}_{\mathrm{m}}=150$ (4.8 hours), and $\mathrm{N}_{\mathrm{m}}=100$ (3.2 hours), respectively, always with the data points selected through SIG. Notably, the 3.2-hour case performs comparably to the reference model (yellow line) while achieving a $60\ \%$ reduction in measurement time. Numerical values corresponding to these plots are provided in Tables~\ref{f1scorelowfield}, \ref{maeAlowfield}, and \ref{maeBlowfield} of Sec.~\ref{numericalresults} in the SM~\cite{supplementalmaterial}.}
\label{reductionmetricsimage}
\end{figure*}

\subsection{Experimental validation of the SIG approach}

We apply the SIG approach to data points within the ranges $\tau_{32}\in[6,\ 50] \ \mathrm{\mu s}$ and $\tau_{256}\in[10,\ 40] \ \mathrm{\mu s}$, with time resolutions $\Delta \tau_{32} = \Delta \tau_{256} = 4\ \mathrm{ns}$. The variance is computed across 50,000 synthetic samples, and the $\mathrm{N}_{\mathrm{p}}=4000$ data points with the highest variance are selected from each signal ($\mathrm{N}=32$ and $\mathrm{N}=256$). This selection leads to a reduction in measurement time of approximately $50\ \%$.
The model is trained using these selected data points and subsequently tested on the experimental data. The resulting output image is the right image of Fig.~\ref{outputimages}~(a). The model predicts 27 nuclear spins, of which 25 match previous predictions (green boxes).
This suggests that, although the model’s performance declines slightly, as shown in all plots of Fig.~\ref{outputimages}~(b), the reduction in the number of detected nuclear spins is minimal --only two fewer than in the full-data case. For many characterization tasks, including those needed to construct a quantum memory, this number remains sufficient~\cite{dobrovitski06, bradley19, jing23}.

Notably, this result is achieved with half the measurement time compared to the full dataset, yet with only a minimal loss in performance: the model detects 25 nuclear spins instead of 27. This demonstrates that the selected data points retain the most informative features of the signal and are sufficient for accurate system characterization.

\subsection{Robustness to increased shot noise}

Another strategy for measurement time reduction is increasing the shot noise in the input signals. In particular, we analyze the cases with  $\mathrm{N}_{\mathrm{m}}=200$, $\mathrm{N}_{\mathrm{m}}=150$, and $\mathrm{N}_{\mathrm{m}}=100$. We  generate the same test set as in the original dataset but modify the number of measurements used to compute the average at each data point, reducing it from $\mathrm{N}_{\mathrm{m}}=250$ to the values specified above. These modified test sets are then evaluated using the model in which the measurement time had already been reduced to 4 hours through the SIG approach. This model was trained using $\mathrm{N}_{\mathrm{m}}=250$ measurements.

In Fig~\ref{reductionmetricsimage}~(a), we observe that for $\mathrm{N}_{\mathrm{m}}=200$ (3.2 hours) and $\mathrm{N}_{\mathrm{m}}=150$ (2.4 hours) the model's performance remains nearly identical to that with $\mathrm{N}_{\mathrm{m}}=250$ (4 hours). In the case of $\mathrm{N}_{\mathrm{m}}=100$ (1.6 hours), the model's performance slightly decreases. Despite this minor degradation, the accuracy of the system’s characterization remains largely unaffected. This measurement-reduction proposal suggests that precise system characterization is possible with just \( \mathrm{N}_{\mathrm{m}} = 100 \) measurements per data point, resulting in significantly faster data acquisition.

\section{Measurement time reduction: low field} \label{experimentaltimelowfield}

In this section, we focus on the low-field regime (\(B_z = 40.4\,\mathrm{G}\)), which is one order of magnitude lower than the high-field case. In this regime, the signal becomes significantly more complex, as \(P_+\) no longer consists of a sequence of dips. In particular, we find that distinct samples containing more than ten nuclear spins can produce very similar signals, making them difficult to distinguish (see Fig.~\ref{lowfieldsignalsimage} in Sec.~\ref{lowfieldsignals} of the SM~\cite{supplementalmaterial}). Despite the increased complexity of the data analysis, this regime is experimentally attractive, as applying and aligning smaller magnetic fields is typically much simpler and more convenient.
In this scenario, we evaluate the methodology introduced in Sec.~\ref{sec:informationgain} only on synthetic data generated from simulations.

To assess the effectiveness of the SIG approach at low magnetic field, we first train the SALI-based model using 2 input signals constructed identically to those in the high-field regime (see Sec.~\ref{modelinginputsignals}); this serves as our reference model. Acquiring these two input signals experimentally would take approximately 8 hours in the lab, taking into account that each data point is measured $\mathrm{N}_{\mathrm{m}}=250$ times.

All other datasets at low field, apart from the reference, are generated from selected $\tau$ values within the range $\tau\in~[1,\ 50] \ \mathrm{\mu s}$, with a resolution of $\Delta \tau = 1\ \mathrm{ns}$.
Improving the resolution of the measurements serves as a counterbalance to the reduced magnetic field. Based on the SIG approach, we select the $\mathrm{N}_{\mathrm{p}}=8000$ data points with the highest variance within the specified $\tau$ range to construct the input signals for pulse sequences with $\mathrm{N} = 32$ and $\mathrm{N} = 256$ $\pi$-pulses. The total measurement time an experimentalist would need to acquire these selected data points in the lab is approximately 8 hours, which matches the time required to measure the input signals used in the reference model. The model trained on this higher-resolution data, selected via SIG, outperforms the reference model while requiring the same measurement time.

In all plots of Fig.~\ref{reductionmetricsimage}~(b), the yellow line represents the reference model, while the black line corresponds to the model incorporating the SIG-based selection of data points. We test the SIG-incorporating model’s resilience to shot noise using synthetic data. Measuring each selected data point $\mathrm{N}_{\mathrm{m}}=100$ times would amount to a total measurement time of about 3.2 hours. Despite this reduced time, Fig.~\ref{reductionmetricsimage}~(b) shows that the performance of the 3.2-hour model is comparable to --or even better than-- that of the reference model, which requires 8 hours. Again, we do not train a different model for each $\mathrm{N}_{\mathrm{m}}$.

Although characterization at low magnetic field remains significantly less effective than at high field, this approach serves as a step toward optimizing data collection strategies to enhance model performance in regimes that were previously thought to be impractical --both in terms of magnetic field strength and the required number of repetitions per decoupling sequence.
By applying the SIG approach and measuring for 8 hours, the model can detect up to 20 nuclear spins, bringing it closer to practical applications in real characterization problems.

\section{Conclusions}

In this work, we have proposed and demonstrated an offline strategy to optimize data acquisition for the characterization of nuclear spin clusters with a quantum sensor, an inherently time-consuming task. Our approach is based on a newly introduced figure of merit for information gain: the surrogate information gain (SIG), a computationally tractable, variance-driven quantity that is robust to experimental imperfections and can be used to reduce experimental time while maintaining high informational value. This optimization framework was implemented using a deep learning signal-to-image (SALI) architecture, chosen for its microsecond scale run-time and adaptability to varying input selections; our approach is, however, very general and could be readily extended to other estimation methods, for example, based on Markov Chain Monte Carlo~\cite{galli25}.
We validated our strategy experimentally in the high-field regime and demonstrated its effectiveness theoretically using synthetic data in the low-field regime. In both regimes, our approach yielded substantial time savings. In the high-field regime, the measurement time --defined as the total experimental time required to measure the signals used by the SALI-based model to characterize the nuclear spin cluster-- was reduced by 85$\%$, from 11 hours to 1.6 hours; in the low-field regime, by 60$\%$, from 8 hours to 3.2 hours.
These results demonstrate that statistics-guided measurement selection can significantly improve the efficiency of nuclear spin characterization.

Our proposed strategy is non-adaptive, i.e. it does not require any real-time adaptation based on previous measurement outcomes. This makes it intrinsically easy to implement, as it does not require complicated real-time electronics. An open question for future studies is whether the data acquisition time can be further minimized by deploying real-time adaptive techniques \cite{gebhart23, joas21, arshad24}. Our parallel work from Ref.~\cite{belliardo25} shows that variational Bayesian inference, while featuring a slower run-time than the SALI-based model discussed here, can compute a posterior probability distribution within minutes and can be applied to input measurements that are not pre-defined but instead selected adaptively in real time. This is very promising for the implementation of real-time adaptive techniques, and future work will investigate if the SIG, or other figures of merit or computational approaches \cite{rainforth24} will enable further reduction of the data acquisition and processing time.

Our results impact quantum technologies in several ways. First, they enable fast characterization of the local environment that leads to decoherence of electron spin qubits. The identification of the nuclear spins in the environment may enable the design of tailored control sequences to minimize decoherence and improve gate fidelities \cite{waeber19}.
Second, nuclear spins are a valuable resource for quantum information processing. Owing to their long coherence times, they have been proposed, and in some cases demonstrated, as auxiliary qubits in quantum networking and computing architectures \cite{cramer16, bourassa20, abobeih22, reiner24, marcks25}. The design of nuclear spin control and readout protocols requires precise knowledge of their hyperfine coupling to the central electron spin. Our techniques offer a path toward scaling these technologies to larger qubit arrays. Finally, our work may have a significant impact on the field of nanoscale nuclear magnetic resonance (nano-NMR). Reducing the sensing time can make nano-NMR more practical and broadly applicable~\cite{budakian24, du24, schwartz19, briegel25, aslam17, aslam23}. In particular, for unstable chemical or biological samples, only a shorter data acquisition time might make the measurements possible. 

Although our focus here has been on the identification of individual nuclear spins with a quantum sensor, the proposed SIG approach is generally valid and could be applied to different experiments in multi-parameter quantum systems. For instance, SIG could be employed in multiphase estimation on a photonic circuit, where the controls are tunable phases in the interferometer~\cite{valeri20, valeri23}. Another potential application is waveform estimation in quantum sensors, which could benefit from SIG in finding optimal controls~\cite{tsang11, gardner25}.

\section{Data availability}
\noindent The experimental data analyzed during this study was provided by M. Abobeih and T. H. Taminiau. The synthetic datasets generated for training the neural network are available from the corresponding author upon reasonable request.

\section{Code availability}
\noindent The code used in this study is available from the corresponding author upon reasonable request.

\section{Acknowledgements}
\noindent The authors acknowledge discussions regarding numerical strategies with Oliver T. Whaites and Carlos Munuera-Javaloy. C.~Bonato, E.~Garrote, and J.~Casanova equally led this work.

\section{Author contributions}
\noindent Numerical analyses were carried out by B.~Varona-Uriarte and F.~Belliardo. Experimental data was provided by M. H. Abobeih and T. H. Taminiau. C. Bonato, E. Garrote, and J. Casanova supervised the work. B.~Varona-Uriarte wrote the manuscript with input from F.~Belliardo, C. Bonato, E. Garrote, and J. Casanova.  All authors reviewed the manuscript and actively revised sections to improve clarity and presentation.

\section{Funding}
\noindent J.~Casanova acknowledges the Agencia Estatal de Investigación via the Modelizado, Optimización, y Esquemas de Magnetometria en Centros de Color project PID2024-161371NB-C22, and the Basque Government under Grant No. IT1470-22. J.~Casanova, E.~Garrote and B.~Varona-Uriarte acknowledge the Quench project that has received funding from the European Union’s Horizon Europe - The EU Research and Innovation Programme under grant agreement No 101135742. C.~Bonato acknowledges support by the Engineering and Physical Sciences Research Council (EP/S000550/1, EP/V053779/1, EP/Z533208/1, EP/Z533191/1, UKRI2706) and the European Innovation Council (QuSPARC, grant agreement 101186889).  This work is also supported by the project 23NRM04 NoQTeS, which has received funding from the European Partnership on Metrology, co-financed from the European Unions Horizon Europe Research and Innovation Programme and by the
Participating States.

\section{Competing interests}
\noindent The authors declare no competing interests.

\section{Additional information}
\noindent Correspondence and requests for materials should be addressed to Borja Varona Uriarte.

\clearpage

\widetext

\begin{center}
\textbf{ \large Supplemental Material: \\ Reducing Sensing Time through Offline Experimental Design for Nuclear Spin Detection}
\end{center}

\renewcommand{\thesection}{S\Roman{section}}

\setcounter{equation}{0}
\setcounter{figure}{0}
\setcounter{table}{0}
\setcounter{section}{0}
\setcounter{page}{1} \makeatletter 

\global\long\def\theequation{S\arabic{equation}}
\global\long\def\thefigure{S\arabic{figure}}
\global\long\def\thetable{S\arabic{table}}
\global\long\def\thesection{S\arabic{section}}

\global\long\def\citenumfont#1{#1}

\section{Description of the system} \label{equationderivation}

We consider a system consisting of an NV sensor coupled to $n$ $^{13}\mathrm{C}$ nearby nuclear spins through hyperfine interactions characterized by coupling vectors $\vec{A}_j=(A_j^{z},A_j^{\bot})$. The dynamics of the system are described by the Hamiltonian

\begin{equation}
H=\sum_{j=1}^n \omega_j \ \hat{\omega}_j \cdot\vec{I}_j + \frac{f(t)}{2}\sigma_z \sum_{j=1}^n \vec{A}_j \cdot \vec{I}_j,
\label{Hi}
\end{equation}

\noindent where the effective nuclear spin frequency vector is given by $\omega_j \ \hat{\omega}_j = \gamma_{\rm{n}} B_z \  \hat{z} + \frac{1}{2} \vec{A}_j$. Here, $\gamma_{\rm{n}} = (2\pi) \times 10.705$ MHz/T is the gyromagnetic ratio of the nuclear spins, and $B_z$ is an external magnetic field aligned along the NV axis ($\hat{z}$), defining the Larmor frequency  $\omega_{\rm{L}}=\gamma_{\rm{n}}B_z$. Microwave (MW) $\pi$-pulses are applied to the NV following the CPMG sequence with inter-pulse spacing $\tau$, and the NV state is readout optically. The modulation function $f(t)=\pm 1$ describes the effect of the applied MW driving. By repeating this sequence $\mathrm{N}_{\mathrm{m}}$ times, the survival probability $P_+$ of the NV's initial state can be computed. In the theoretical limit of infinite measurements (note our simulations consider shot noise and decoherence effects), $P_+$ takes the form:

\begin{equation}
P_+ = \frac{1}{2} \left(1+\displaystyle\prod_{j=1}^n M_j\right),
\label{probability}
\end{equation}
where 
\begin{equation}
M_j = 1-m_{j,x}^2 \frac{(1-\cos{\alpha_j})(1-\cos{\beta})}{1+\cos{\alpha_j}\cos{\beta}-m_{j,z}\sin{\alpha_j}\sin{\beta}}\sin{\frac{N\phi_j}{2}}^2 ,
\label{mj}
\end{equation}

\begin{equation}
\cos{\phi_j} = \cos{\alpha_j}\cos{\beta}-m_{j,z}\sin{\alpha_j}\sin{\beta},
\label{cos}
\end{equation} 

\noindent with $m_{j,z} = \frac{(A_j^{z}+\omega_L)}{\tilde{\omega}_j}$, $m_{j,x} = \frac{A_j^{\bot}}{\tilde{\omega}_j}$, $\tilde{\omega}_j = \sqrt{(A_j^{z}+\omega_L)^2+{A_j^{\bot}}^2}$, $\alpha_j = \tilde{\omega}_j\tau$, $\beta=\omega_L \tau$. See the full derivation in Ref.~\cite{staminiau12}.

\section{Training of the models} \label{straining}

The SALI architecture is detailed in the Supplemental Material of
Letter \cite{svarona24}, with the following modifications. The number of filters in the 1-dimensional and 2-dimensional convolutional layers has been increased from 16 and 32 to 32 and 64, respectively. Additionally, the input data normalization has been replaced by a batch normalization layer at the network's input.

All models in this study were trained on 5 million samples, divided into 70 \% for training, 15 \% for validation, and 15 \% for testing. Separate models were trained for each data point selection: the 2-input model, the 4-input model, and the SIG-incorporating model at high magnetic field, as well as the reference model and the SIG-incorporating model at low magnetic field. The network was trained using the Mean Square Error (MSE) loss function, calculated across all pixels in the output grid, and optimization was performed with the Adam optimizer, initialized with a learning rate of $1 \times 10^{-3}$. The learning rate was reduced during training, with a patience of 2 and a learning rate factor of 0.1. The neural network was implemented in PyTorch and trained for approximately two days on a single NVIDIA H100 80GB GPU.

\section{Nuclear spin predictions} \label{numericalpredictions}

\begin{table}[h]
\centering
\renewcommand{\arraystretch}{1.75}
\begin{tabular}{ |w{c}{1.8cm}|w{c}{1.8cm}|w{c}{1.8cm}|w{c}{1.8cm}|w{c}{1.8cm}|w{c}{1.8cm}|w{c}{1.8cm}| }
\cline{2-7}
\multicolumn{1}{c|}{} & \multicolumn{2}{c|}{\textbf{2 inputs}} & \multicolumn{2}{c|}{\textbf{4 inputs}} & \multicolumn{2}{c|}{\textbf{SIG}} \\ 
\hline
\textbf{Nucleus} & $A^{z}$ (kHz) & $A^{\bot}$ (kHz) & $A^{z}$ (kHz) & $A^{\bot}$ (kHz) & $A^{z}$ (kHz) & $A^{\bot}$ (kHz) \\ 

\hline
C1 & - & - & - & - & - & - \\
C2 & -24.32(0) & 22.07(1) & -24.28(0) & 22.11(1) & -24.33(1) & 22.10(2) \\
C3 & -20.64(0) & 41.54(1) & -20.66(0) & 41.66(3) & -20.63(1) & 41.50(2) \\
C4 & -17.75(1) & 8.73(2) & -17.71(1) & 9.00(2) & -17.64(1) & 8.84(5) \\
C5 & -14.8(1) & 4.7(7) & -14.67(1) & 9.29(1) & -14.56(1) & 10.40(5) \\
C6 & -7.96(1) & 15.60(3) & -7.93(1) & 5.11(4) & -7.97(2) & 19.30(3) \\
C7 & -8.07(0) & 20.73(1) & -8.14(1) & 21.30(2) & -8.02(1) & 20.60(4) \\
C8 & -7.86(1) & 6.05(3) & -0.86(2) & 6.48(4) & -5.28(1) & 12.80(4) \\
C9 & -4.79(1) & 2.83(2) & -4.82(1) & 3.59(1) & -5.19(3) & 2.75(3) \\
C10 & -2.93(0) & 10.13(1) & -2.79(1) & 7.93(2) & -3.05(0) & 12.60(3) \\
C11 - C16 & -2.54(0) & 12.60(1) & -1.09(3) & 12.80(5) & - & - \\
C17 - C22 & -1.04(0) & 12.96(1) & -1.13(1) & 13.6(1) & - & - \\
C23 - C27 & 1.94(0) & 13.11(1) & 1.93(0) & 13.19(1) & - & - \\
C28 - C29 & 1.95(0) & 13.06(1) & 2.29(2) & 4.79(4) & 13.85(1) & 8.9(1) \\
C30 & 2.53(3) & 9.55(1) & 3.51(1) & 5.03(1) & 14.01(3) & 6.3(1) \\
C31-32 & 4.03(1) & 8.15(2) & 4.30(0) & 7.29(2) & 4.19(1) & 8.11(3) \\
C33 & 4.07(1) & 9.19(1) & - & - & 3.77(3) & 10.80(4) \\
C34 & 4.26(0) & 12.16(4) & - & - & 3.90(3) & 12.6(2) \\
C35 & 7.04(1) & 2.13(1) & 7.01(0) & 2.61(2) & 7.00(1) & 2.27(1) \\
C36 & - & - & - & - & 14.1(1) & 5.5(1) \\
C37 & 9.19(1) & 5.11(4) & 9.17(1) & 5.14(3) & 9.24(1) & 5.16(3) \\
C38 & 13.82(0) & 7.97(6) & 13.99(0) & 8.7(1) & 13.80(1) & 7.90(3) \\
C39* & - & - & 14.06(2) & 6.85(2) & 13.73(2) & 16.23(8) \\
C40 & 11.38(0) & 58.82(1) & 11.29(0) & 58.75(2) & 11.37(1) & 59.17(2) \\
C41 - C42 & 19.84(1) & 5.71(2) & 20.01(1) & 5.96(1) & 19.84(1) & 5.98(3) \\
C43* & - & - & 23.18(2) & 6.70(6) & - & - \\
C44* & 27.87(5) & 2.91(7) & 28.55(3) & 3.03(3) & - & - \\
C45 & 32.86(5) & 3.00(6) & 31.46(5) & 3.26(7) & 32.38(3) & 2.59(4) \\
C46* & 44.30(2) & 3.81(5) & 44.19(1) & 3.65(3) & 48.28(2) & 12.23(4) \\
C47 & 36.23(1) & 26.44(3) & 36.22(1) & 26.48(2) & 36.36(1) & 26.55(2) \\
C48 & 48.33(1) & 11.12(3) & 48.50(0) & 10.17(2) & 48.36(1) & 11.03(6) \\
\hline
\multicolumn{1}{|c|}{\textbf{Total:}} & \multicolumn{2}{c|}{\textbf{27}} & \multicolumn{2}{c|}{\textbf{27}} &  \multicolumn{2}{c|}{\textbf{25}}  \\
\hline
\end{tabular}
\caption{Numerical values of the predicted coupling constants for the spins highlighted by green boxes in Fig.~\ref{outputimages}~(a). These predictions match those found in Supplementary Table 2 of Letter~\cite{sjung21}.}
\label{predictions}
\end{table}

\begin{table}[h]
\centering
\renewcommand{\arraystretch}{1.4}
\begin{tabular}{ |w{c}{1.8cm}|w{c}{1.8cm}|w{c}{1.8cm}|w{c}{1.8cm}|w{c}{1.8cm}|w{c}{1.8cm}|w{c}{1.8cm}| }
\cline{2-7}
\multicolumn{1}{c|}{} & \multicolumn{2}{c|}{\textbf{2 inputs}} & \multicolumn{2}{c|}{\textbf{4 inputs}} & \multicolumn{2}{c|}{\textbf{SIG}} \\ 
\hline
\textbf{Nucleus} & $A^{z}$ (kHz) & $A^{\bot}$ (kHz) & $A^{z}$ (kHz) & $A^{\bot}$ (kHz) & $A^{z}$ (kHz) & $A^{\bot}$ (kHz) \\ 

\hline
C1E & -43.20(4) & 2.71(3) & -42.78(2) & 3.24(2) & -42.91(3) & 2.78(3) \\
C2E & - & - & -30.93(6) & 2.43(4) & -30.12(2) & 2.59(1) \\
C3E & -26.31(2) & 3.14(4) & -26.91(6) & 3.3(1) & - & - \\
\hline
\multicolumn{1}{|c|}{\textbf{Total:}} & \multicolumn{2}{c|}{\textbf{2}} & \multicolumn{2}{c|}{\textbf{3}} &  \multicolumn{2}{c|}{\textbf{2}}  \\
\hline
\end{tabular}
\caption{Numerical values of the predicted coupling constants for the spins highlighted by red boxes in Fig.~\ref{outputimages}~(a).
These predictions do not match those found in Supplementary Table 2 of Letter~\cite{sjung21}. The strongly coupled nuclear spin (C1 in Table~\ref{predictions}) reported in that work lies outside our sampling range and is therefore not detected in our predictions. Instead, our model identifies these other spins within our sampling range that may be influenced by C1.}
\label{predictionsnomatch}
\end{table}

To evaluate the reliability of the predicted nuclear spins, a small Gaussian noise (mean = 0, standard deviation = 0.005) was added to the experimental signals. This simulates classical noise sources present in real measurements, beyond the shot noise already accounted for in the simulations of the $P_+$ signals. Each model configuration (2 inputs, 4 inputs, and SIG) was tested 10 times, with different random noise added in each run. The mean values and standard deviations of the predicted coupling constants --corresponding to the images shown in Fig.~\ref{outputimages}~(a) of the main text-- were computed across these 10 runs and are reported in Tables~\ref{predictions} and~\ref{predictionsnomatch}.

\section{Definitions of the metrics} \label{metricsdescription}

To evaluate the performance of the deep learning model in detecting the number of nuclear spins, we use standard classification metrics: precision (P), recall (R), and the F1-score. These metrics are based on three types of outcomes when comparing predicted spins to the ground truth. A true positive (TP) is a predicted spin that correctly matches a ground-truth spin. A false positive (FP) is a predicted spin that does not correspond to any ground-truth spin. A false negative (FN) is a ground-truth spin that the model failed to detect. Precision measures the fraction of predicted spins that are correct, while recall measures the fraction of ground-truth spins that are successfully detected. The F1-score is the harmonic mean of precision and recall, providing a single metric that balances both false positives and false negatives. The mathematical definitions of the three metrics are:

\begin{equation}
\mathrm{P} = \frac{\mathrm{TP}}{\mathrm{TP} + \mathrm{FP}}, \quad
\mathrm{R} = \frac{\mathrm{TP}}{\mathrm{TP} + \mathrm{FN}}, \quad
\mathrm{F1-score} = 2 \cdot \frac{\mathrm{P} \cdot \mathrm{R}}{\mathrm{P} + \mathrm{R}}.
\end{equation}

In addition to evaluating the detection of the number of nuclear spins, we also assess the accuracy of parameter estimation by computing the Mean Absolute Error (MAE) of the predicted coupling constants. This metric quantifies the average absolute deviation between the predicted values $\vec{A}_j^{\,\,\mathrm{pred}}$ and the corresponding ground-truth values $\vec{A}_j^{\,\,\mathrm{true}}$, computed across all correctly identified spins (true positives):

\begin{equation}
\mathrm{MAE}=\frac{1}{n} \sum_{j=1}^n \left|\vec{A}_j^{\,\,\mathrm{true}}-\vec{A}_j^{\,\,\mathrm{pred}}\right|.
\label{maeequation}
\end{equation}
All metrics are computed across all nuclear spins within the test samples, and the plots in Fig.~\ref{outputimages} (b) and Fig.~\ref{reductionmetricsimage} in the main text show the average values for samples containing from 1 to 50 nuclear spins.

\newpage 

\section{Low-field regime signals} \label{lowfieldsignals}
 
\begin{figure*}[h]
\includegraphics[width=1\linewidth]{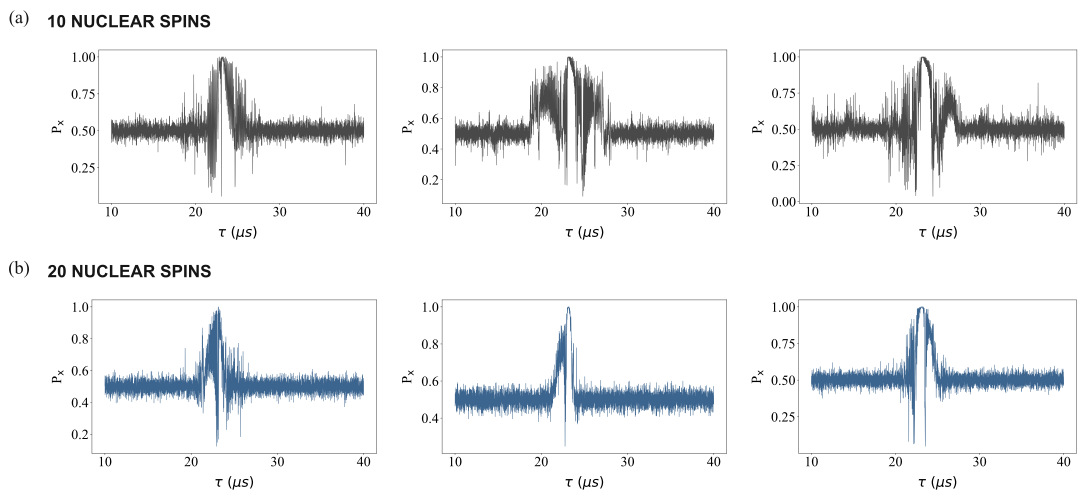}
\caption{(a) Comparison of signals from three different samples, each with 10 nuclear spins coupled to the NV sensor. With this number of spins, the signals become similar, differing only in a small region. Distinct data points are the only ones useful for distinguishing between samples, while regions where all signals share similar value do not contribute to differentiation. (b) Comparison of signals from three different samples with 20 nuclear spins. In this case, the signals exhibit even greater similarity than with 10 nuclear spins. The number of distinct data points between samples is further reduced.}
\label{lowfieldsignalsimage}
\end{figure*}

\newpage 

\section{Numerical results from the main text: metrics} \label{numericalresults}

\begin{table}[h]
\renewcommand{\arraystretch}{1.04}
\begin{tabular}{ |w{c}{2cm}|w{c}{1.8cm}|w{c}{1.8cm}|w{c}{1.8cm}|w{c}{1.8cm}|w{c}{1.8cm}|w{c}{1.8cm}| }
\cline{2-7}
\multicolumn{1}{c|}{} & \multicolumn{6}{c|}{\textbf{F1-score: high-field regime}} \\ 
\hline
\textbf{Nuclear spins} & \textbf{11 hours} & \textbf{8 hours} & \textbf{4 hours} & \textbf{3.2 hours} & \textbf{2.4 hours} & \textbf{1.6 hours} \\
\hline
1 & 0.6486 & 0.8638 & 0.9579 & 0.9476 & 0.9292 & 0.8951 \\
2 & 0.8997 & 0.9583 & 0.9695 & 0.9690 & 0.9676 & 0.9625 \\
3 & 0.9724 & 0.9820 & 0.9685 & 0.9682 & 0.9677 & 0.9658 \\
4 & 0.9899 & 0.9869 & 0.9663 & 0.9659 & 0.9654 & 0.9634 \\
5 & 0.9920 & 0.9867 & 0.9629 & 0.9625 & 0.9617 & 0.9596 \\
6 & 0.9920 & 0.9849 & 0.9612 & 0.9609 & 0.9598 & 0.9572 \\
7 & 0.9912 & 0.9830 & 0.9578 & 0.9571 & 0.9561 & 0.9527 \\
8 & 0.9899 & 0.9810 & 0.9543 & 0.9537 & 0.9523 & 0.9484 \\
9 & 0.9882 & 0.9785 & 0.9520 & 0.9514 & 0.9494 & 0.9452 \\
10 & 0.9875 & 0.9766 & 0.9481 & 0.9474 & 0.9452 & 0.9403 \\
11 & 0.9859 & 0.9735 & 0.9446 & 0.9438 & 0.9414 & 0.9356 \\
12 & 0.9845 & 0.9711 & 0.9414 & 0.9401 & 0.9376 & 0.9315 \\
13 & 0.9833 & 0.9692 & 0.9381 & 0.9366 & 0.9338 & 0.9267 \\
14 & 0.9810 & 0.9654 & 0.9338 & 0.9321 & 0.9289 & 0.9207 \\
15 & 0.9797 & 0.9626 & 0.9304 & 0.9288 & 0.9248 & 0.9157 \\
16 & 0.9780 & 0.9598 & 0.9262 & 0.9236 & 0.9198 & 0.9096 \\
17 & 0.9760 & 0.9571 & 0.9226 & 0.9204 & 0.9159 & 0.9049 \\
18 & 0.9743 & 0.9538 & 0.9180 & 0.9154 & 0.9104 & 0.8978 \\
19 & 0.9726 & 0.9506 & 0.9142 & 0.9115 & 0.9062 & 0.8926 \\
20 & 0.9705 & 0.9475 & 0.9102 & 0.9074 & 0.9012 & 0.8865 \\
21 & 0.9690 & 0.9439 & 0.9057 & 0.9024 & 0.8962 & 0.8799 \\
22 & 0.9665 & 0.9405 & 0.9009 & 0.8976 & 0.8906 & 0.8738 \\
23 & 0.9642 & 0.9373 & 0.8961 & 0.8926 & 0.8850 & 0.8667 \\
24 & 0.9620 & 0.9332 & 0.8917 & 0.8878 & 0.8796 & 0.8596 \\
25 & 0.9596 & 0.9296 & 0.8869 & 0.8829 & 0.8738 & 0.8519 \\
26 & 0.9567 & 0.9257 & 0.8816 & 0.8769 & 0.8675 & 0.8440 \\
27 & 0.9544 & 0.9216 & 0.8764 & 0.8719 & 0.8619 & 0.8360 \\
28 & 0.9519 & 0.9172 & 0.8714 & 0.8665 & 0.8553 & 0.8277 \\
29 & 0.9493 & 0.9131 & 0.8658 & 0.8598 & 0.8487 & 0.8195 \\
30 & 0.9463 & 0.9086 & 0.8602 & 0.8546 & 0.8416 & 0.8111 \\
31 & 0.9428 & 0.9039 & 0.8545 & 0.8480 & 0.8349 & 0.8020 \\
32 & 0.9404 & 0.8999 & 0.8491 & 0.8419 & 0.8284 & 0.7926 \\
33 & 0.9371 & 0.8950 & 0.8433 & 0.8357 & 0.8214 & 0.7842 \\
34 & 0.9340 & 0.8903 & 0.8370 & 0.8289 & 0.8131 & 0.7737 \\
35 & 0.9304 & 0.8851 & 0.8312 & 0.8228 & 0.8058 & 0.7639 \\
36 & 0.9276 & 0.8805 & 0.8258 & 0.8170 & 0.7985 & 0.7550 \\
37 & 0.9237 & 0.8756 & 0.8197 & 0.8102 & 0.7915 & 0.7458 \\
38 & 0.9205 & 0.8702 & 0.8127 & 0.8033 & 0.7838 & 0.7350 \\
39 & 0.9172 & 0.8649 & 0.8073 & 0.7969 & 0.7759 & 0.7257 \\
40 & 0.9130 & 0.8593 & 0.7993 & 0.7890 & 0.7669 & 0.7139 \\
41 & 0.9088 & 0.8538 & 0.7937 & 0.7827 & 0.7590 & 0.7037 \\
42 & 0.9045 & 0.8485 & 0.7868 & 0.7748 & 0.7511 & 0.6928 \\
43 & 0.9009 & 0.8425 & 0.7800 & 0.7672 & 0.7420 & 0.6824 \\
44 & 0.8965 & 0.8364 & 0.7727 & 0.7592 & 0.7339 & 0.6730 \\
45 & 0.8922 & 0.8300 & 0.7651 & 0.7516 & 0.7250 & 0.6610 \\
46 & 0.8887 & 0.8248 & 0.7585 & 0.7442 & 0.7164 & 0.6510 \\
47 & 0.8837 & 0.8182 & 0.7506 & 0.7363 & 0.7069 & 0.6396 \\
48 & 0.8787 & 0.8111 & 0.7425 & 0.7275 & 0.6971 & 0.6276 \\
49 & 0.8747 & 0.8052 & 0.7350 & 0.7190 & 0.6882 & 0.6174 \\
50 & 0.8694 & 0.7983 & 0.7268 & 0.7107 & 0.6783 & 0.6057 \\
\hline
\end{tabular}
\caption{Numerical results from the left plots of Fig.~\ref{outputimages}~(b) and Fig.~\ref{reductionmetricsimage}~(a) in the main text.}
\label{f1scorehighfield}
\end{table}

\begin{table}[h]
\renewcommand{\arraystretch}{1.2}
\begin{tabular}{ |w{c}{2cm}|w{c}{1.8cm}|w{c}{1.8cm}|w{c}{1.8cm}|w{c}{1.8cm}|w{c}{1.8cm}|w{c}{1.8cm}| }
\cline{2-7}
\multicolumn{1}{c|}{} & \multicolumn{6}{c|}{\textbf{MAE of $A^z$ (Hz): high-field regime}} \\ 
\hline
\textbf{Nuclear spins} & \textbf{11 hours} & \textbf{8 hours} & \textbf{4 hours} & \textbf{3.2 hours} & \textbf{2.4 hours} & \textbf{1.6 hours} \\
\hline
1 & 37.3579 & 55.9494 & 34.1342 & 34.4847 & 35.1491 & 36.7516 \\
2 & 30.5994 & 36.5999 & 32.3540 & 32.5773 & 32.7347 & 33.5186 \\
3 & 29.6355 & 31.6862 & 34.2294 & 33.9646 & 34.2671 & 35.0037 \\
4 & 30.4830 & 31.5786 & 35.6287 & 35.6959 & 36.2522 & 36.7835 \\
5 & 32.2229 & 33.2231 & 37.7326 & 37.9817 & 38.2145 & 39.2471 \\
6 & 33.3882 & 34.8250 & 39.4208 & 39.6920 & 39.9132 & 41.0156 \\
7 & 35.0246 & 36.8593 & 41.4483 & 41.6264 & 42.1264 & 43.0197 \\
8 & 36.9964 & 38.9334 & 43.0666 & 43.3798 & 44.0352 & 44.9285 \\
9 & 38.4091 & 40.7201 & 45.0028 & 45.5925 & 46.1187 & 47.2475 \\
10 & 40.0213 & 42.5951 & 46.9020 & 47.5271 & 47.8071 & 49.2318 \\
11 & 41.8209 & 44.6581 & 48.9746 & 49.6040 & 49.9161 & 51.6924 \\
12 & 43.0590 & 46.2422 & 50.6468 & 50.9285 & 51.8300 & 53.2450 \\
13 & 44.7005 & 48.1559 & 52.3590 & 52.9511 & 53.7095 & 55.3132 \\
14 & 46.4185 & 50.0166 & 54.2082 & 54.9139 & 55.6775 & 57.2206 \\
15 & 47.8875 & 51.8365 & 56.3029 & 56.6690 & 57.5752 & 59.4111 \\
16 & 49.9397 & 53.8908 & 58.4786 & 58.7619 & 59.6359 & 61.2832 \\
17 & 51.0567 & 55.3414 & 59.7513 & 60.1687 & 61.3133 & 63.0093 \\
18 & 52.7545 & 57.2409 & 61.5724 & 62.2434 & 63.1865 & 65.0949 \\
19 & 54.0236 & 59.0522 & 63.5470 & 64.2396 & 65.0678 & 66.7282 \\
20 & 55.8557 & 60.8222 & 65.5818 & 66.0567 & 66.9224 & 69.1059 \\
21 & 57.5917 & 62.6874 & 67.3422 & 67.8250 & 68.8935 & 71.2285 \\
22 & 59.2823 & 64.6295 & 69.4162 & 70.0028 & 71.2196 & 73.4623 \\
23 & 60.4226 & 66.4661 & 71.2642 & 72.0325 & 72.9794 & 75.2506 \\
24 & 62.6901 & 68.5863 & 73.3250 & 74.0075 & 75.1410 & 77.5501 \\
25 & 64.4751 & 70.7731 & 75.7269 & 76.2446 & 77.5201 & 79.8323 \\
26 & 66.0593 & 72.5829 & 77.8761 & 78.4538 & 79.6114 & 81.6559 \\
27 & 67.5344 & 74.4330 & 79.2851 & 80.1474 & 81.0705 & 83.4549 \\
28 & 69.6870 & 76.6909 & 81.6278 & 82.3435 & 83.4338 & 85.9400 \\
29 & 71.4118 & 78.6176 & 83.9669 & 84.7580 & 85.7953 & 87.9928 \\
30 & 73.4251 & 80.8052 & 86.3702 & 86.9349 & 88.1992 & 90.5276 \\
31 & 75.5411 & 83.0562 & 88.3005 & 89.4740 & 90.3810 & 92.5552 \\
32 & 77.2835 & 84.9648 & 90.8000 & 91.2820 & 92.4126 & 95.2764 \\
33 & 79.4161 & 87.2626 & 93.1314 & 93.9148 & 95.0340 & 97.2793 \\
34 & 81.5946 & 89.7932 & 95.5567 & 95.9737 & 97.3636 & 99.8144 \\
35 & 83.5816 & 91.9698 & 97.9221 & 98.9156 & 99.9614 & 102.1157 \\
36 & 85.8841 & 93.7312 & 100.4024 & 101.1566 & 101.9878 & 104.5278 \\
37 & 87.9473 & 96.1841 & 103.3925 & 103.5926 & 104.4820 & 107.1818 \\
38 & 90.1387 & 98.5522 & 105.8394 & 106.3856 & 107.4910 & 110.0982 \\
39 & 92.4781 & 101.1007 & 108.2088 & 108.9615 & 109.9228 & 112.9811 \\
40 & 94.7375 & 103.7605 & 111.3539 & 112.3530 & 112.5630 & 115.6889 \\
41 & 97.2296 & 106.7776 & 114.6005 & 114.8631 & 116.2304 & 118.9873 \\
42 & 99.4325 & 108.7256 & 116.5325 & 117.1347 & 118.7316 & 121.0262 \\
43 & 102.1366 & 111.5452 & 119.7913 & 120.5075 & 121.1770 & 124.3705 \\
44 & 104.2050 & 114.0404 & 122.7284 & 123.2835 & 123.7517 & 127.0794 \\
45 & 107.0253 & 116.8003 & 125.2068 & 126.1152 & 127.1110 & 129.4451 \\
46 & 109.4121 & 119.3892 & 128.4924 & 129.0681 & 130.4413 & 133.5998 \\
47 & 112.2885 & 122.0577 & 131.2663 & 131.8696 & 133.2541 & 136.4695 \\
48 & 114.3209 & 124.9334 & 134.5305 & 135.3156 & 136.1304 & 139.7813 \\
49 & 116.9255 & 127.3942 & 137.2424 & 137.4102 & 138.9628 & 143.2425 \\
50 & 119.7766 & 130.8862 & 140.3613 & 141.2116 & 142.8719 & 145.9034 \\
\hline
\end{tabular}
\caption{Numerical results from the middle plots of Fig.~\ref{outputimages}~(b) and Fig.~\ref{reductionmetricsimage}~(a) in the main text.}
\label{maeAhighfield}
\end{table}

\begin{table}[h]
\renewcommand{\arraystretch}{1.2}
\begin{tabular}{ |w{c}{2cm}|w{c}{1.8cm}|w{c}{1.8cm}|w{c}{1.8cm}|w{c}{1.8cm}|w{c}{1.8cm}|w{c}{1.8cm}| }
\cline{2-7}
\multicolumn{1}{c|}{} & \multicolumn{6}{c|}{\textbf{MAE of $A^{\bot}$ (Hz): high-field regime}} \\ 
\hline
\textbf{Nuclear spins} & \textbf{11 hours} & \textbf{8 hours} & \textbf{4 hours} & \textbf{3.2 hours} & \textbf{2.4 hours} & \textbf{1.6 hours} \\
\hline
1 & 92.7411 & 179.0140 & 99.5267 & 101.9721 & 105.8136 & 111.5539 \\
2 & 82.9324 & 118.2944 & 96.1301 & 98.1266 & 101.8063 & 109.6251 \\
3 & 84.3050 & 103.4392 & 103.1559 & 106.2852 & 109.4395 & 116.2342 \\
4 & 87.9073 & 105.2502 & 111.9473 & 114.4902 & 118.2498 & 124.2189 \\
5 & 94.2537 & 113.0375 & 121.6607 & 122.8852 & 126.3362 & 133.9693 \\
6 & 100.5010 & 120.1336 & 129.0031 & 131.0350 & 135.1029 & 142.0987 \\
7 & 107.8701 & 128.9713 & 138.7053 & 140.3192 & 144.9082 & 150.3535 \\
8 & 114.3327 & 137.0794 & 145.5136 & 147.8620 & 151.7646 & 159.0364 \\
9 & 121.7272 & 144.6708 & 154.7982 & 157.4919 & 160.7810 & 168.6182 \\
10 & 128.7349 & 152.9570 & 163.0010 & 166.2905 & 169.3529 & 176.4748 \\
11 & 136.9179 & 161.5221 & 172.1821 & 175.3291 & 178.7967 & 186.0942 \\
12 & 143.0285 & 169.3807 & 180.2560 & 182.1540 & 186.8779 & 194.2016 \\
13 & 149.5902 & 177.9830 & 187.9096 & 191.0126 & 195.5303 & 202.0678 \\
14 & 158.0408 & 186.6083 & 198.2371 & 201.7923 & 205.0219 & 212.4405 \\
15 & 165.2590 & 194.2375 & 206.5662 & 209.2172 & 213.1823 & 219.8407 \\
16 & 173.0454 & 203.0749 & 215.0816 & 217.4856 & 220.3072 & 227.5742 \\
17 & 179.3371 & 210.5368 & 222.3453 & 225.6316 & 229.3309 & 235.3441 \\
18 & 188.2014 & 220.5495 & 231.9111 & 233.6932 & 238.3172 & 245.1766 \\
19 & 194.0988 & 227.8202 & 239.8604 & 242.7204 & 246.4353 & 253.1864 \\
20 & 203.6118 & 237.2023 & 249.2740 & 251.8718 & 255.4112 & 262.5605 \\
21 & 211.7907 & 246.0937 & 258.6246 & 261.3748 & 264.0440 & 270.4297 \\
22 & 219.8975 & 255.2739 & 267.3053 & 269.8192 & 273.3937 & 280.7072 \\
23 & 226.8004 & 263.7501 & 275.4572 & 278.6257 & 281.6886 & 289.8051 \\
24 & 236.0165 & 272.0377 & 284.8483 & 288.0743 & 289.9856 & 297.9279 \\
25 & 243.0889 & 282.5232 & 295.4275 & 296.4088 & 300.8389 & 307.0539 \\
26 & 253.7127 & 292.7961 & 304.4774 & 307.5744 & 310.0954 & 315.6405 \\
27 & 261.1343 & 299.7650 & 311.2469 & 314.6049 & 317.6979 & 322.6385 \\
28 & 269.4547 & 309.1641 & 322.4268 & 324.7590 & 327.2547 & 332.9363 \\
29 & 277.9067 & 317.9956 & 329.7712 & 332.4295 & 334.8528 & 340.2266 \\
30 & 287.1657 & 328.5346 & 341.5482 & 343.8064 & 345.3277 & 350.5673 \\
31 & 297.5729 & 337.9642 & 350.3327 & 352.3903 & 353.8502 & 359.0556 \\
32 & 305.3029 & 345.7488 & 359.8491 & 360.2178 & 363.1237 & 367.4391 \\
33 & 314.8282 & 356.6980 & 369.6890 & 369.8393 & 372.4225 & 376.7413 \\
34 & 324.4652 & 366.3151 & 378.4167 & 378.2029 & 380.8598 & 385.7413 \\
35 & 334.4574 & 375.3354 & 388.8010 & 390.9910 & 391.2362 & 393.9228 \\
36 & 343.5272 & 383.6141 & 397.1437 & 397.9214 & 399.2224 & 401.6898 \\
37 & 351.7478 & 393.1965 & 406.7028 & 407.1157 & 407.5269 & 411.6026 \\
38 & 362.5028 & 403.1014 & 416.3320 & 418.3173 & 418.5770 & 421.3554 \\
39 & 373.2334 & 413.8495 & 427.0818 & 428.5969 & 428.1623 & 432.5115 \\
40 & 382.9450 & 424.6684 & 437.2881 & 438.6393 & 438.8290 & 440.5673 \\
41 & 392.6071 & 433.4896 & 447.0351 & 447.6415 & 448.4028 & 450.5998 \\
42 & 402.8465 & 441.9363 & 456.4540 & 456.7807 & 457.7534 & 459.2062 \\
43 & 412.7161 & 452.6886 & 467.2235 & 467.2280 & 467.0481 & 469.6938 \\
44 & 422.7184 & 460.8041 & 477.1734 & 476.2853 & 476.6146 & 477.7239 \\
45 & 433.5034 & 472.0808 & 485.6310 & 486.0177 & 485.4702 & 487.5539 \\
46 & 442.3113 & 481.4498 & 494.7578 & 496.0935 & 495.9943 & 498.6996 \\
47 & 454.4903 & 491.1692 & 505.0525 & 505.4172 & 505.1486 & 507.9029 \\
48 & 462.6673 & 500.6318 & 516.0816 & 514.7325 & 515.4032 & 517.5349 \\
49 & 472.1505 & 509.1696 & 523.0532 & 523.6033 & 523.9010 & 528.0483 \\
50 & 482.9520 & 520.4733 & 533.1283 & 533.3581 & 534.1450 & 537.9038 \\
\hline
\end{tabular}
\caption{Numerical results from the right plots of Fig.~\ref{outputimages}~(b) and Fig.~\ref{reductionmetricsimage}~(a) in the main text.}
\label{maeBhighfield}
\end{table}

\begin{table}[h]
\renewcommand{\arraystretch}{1.2}
\begin{tabular}{ |w{c}{2cm}|w{c}{1.8cm}|w{c}{1.8cm}|w{c}{1.8cm}|w{c}{1.8cm}|w{c}{1.8cm}| }
\cline{2-6}
\multicolumn{1}{c|}{} & \multicolumn{5}{c|}{\textbf{F1-score: low-field regime}} \\ 
\hline
\textbf{Nuclear spins} & \textbf{8 hours} & \textbf{8 hours} & \textbf{6.4 hours} & \textbf{4.8 hours} & \textbf{3.2 hours} \\
\hline
1 & 0.8374 & 0.6058 & 0.5973 & 0.6034 & 0.6214 \\
2 & 0.8461 & 0.8439 & 0.8437 & 0.8456 & 0.8437 \\
3 & 0.8523 & 0.9043 & 0.9008 & 0.9001 & 0.8887 \\
4 & 0.8354 & 0.9085 & 0.9007 & 0.8973 & 0.8775 \\
5 & 0.8035 & 0.8914 & 0.8786 & 0.8741 & 0.8474 \\
6 & 0.7648 & 0.8669 & 0.8499 & 0.8426 & 0.8068 \\
7 & 0.7249 & 0.8427 & 0.8200 & 0.8068 & 0.7648 \\
8 & 0.6855 & 0.8178 & 0.7869 & 0.7705 & 0.7193 \\
9 & 0.6434 & 0.7918 & 0.7529 & 0.7330 & 0.6745 \\
10 & 0.6064 & 0.7648 & 0.7209 & 0.6958 & 0.6313 \\
11 & 0.5674 & 0.7380 & 0.6865 & 0.6599 & 0.5878 \\
12 & 0.5325 & 0.7136 & 0.6573 & 0.6250 & 0.5503 \\
13 & 0.4993 & 0.6901 & 0.6279 & 0.5934 & 0.5163 \\
14 & 0.4722 & 0.6679 & 0.6007 & 0.5622 & 0.4820 \\
15 & 0.4470 & 0.6473 & 0.5752 & 0.5341 & 0.4534 \\
16 & 0.4246 & 0.6272 & 0.5507 & 0.5102 & 0.4271 \\
17 & 0.4036 & 0.6079 & 0.5288 & 0.4851 & 0.4031 \\
18 & 0.3821 & 0.5898 & 0.5074 & 0.4617 & 0.3826 \\
19 & 0.3648 & 0.5706 & 0.4864 & 0.4425 & 0.3626 \\
20 & 0.3483 & 0.5539 & 0.4679 & 0.4237 & 0.3442 \\
21 & 0.3311 & 0.5380 & 0.4507 & 0.4071 & 0.3273 \\
22 & 0.3163 & 0.5226 & 0.4349 & 0.3901 & 0.3137 \\
23 & 0.3034 & 0.5074 & 0.4195 & 0.3755 & 0.2991 \\
24 & 0.2896 & 0.4932 & 0.4054 & 0.3607 & 0.2882 \\
25 & 0.2754 & 0.4782 & 0.3915 & 0.3488 & 0.2759 \\
26 & 0.2643 & 0.4654 & 0.3797 & 0.3372 & 0.2663 \\
27 & 0.2528 & 0.4523 & 0.3676 & 0.3243 & 0.2557 \\
28 & 0.2420 & 0.4411 & 0.3570 & 0.3153 & 0.2478 \\
29 & 0.2313 & 0.4273 & 0.3469 & 0.3050 & 0.2384 \\
30 & 0.2213 & 0.4151 & 0.3373 & 0.2974 & 0.2327 \\
31 & 0.2134 & 0.4033 & 0.3273 & 0.2880 & 0.2252 \\
32 & 0.2035 & 0.3928 & 0.3185 & 0.2816 & 0.2200 \\
33 & 0.1955 & 0.3826 & 0.3104 & 0.2735 & 0.2135 \\
34 & 0.1886 & 0.3717 & 0.3027 & 0.2656 & 0.2070 \\
35 & 0.1818 & 0.3620 & 0.2950 & 0.2591 & 0.2030 \\
36 & 0.1736 & 0.3522 & 0.2880 & 0.2529 & 0.1980 \\
37 & 0.1677 & 0.3441 & 0.2821 & 0.2487 & 0.1939 \\
38 & 0.1628 & 0.3349 & 0.2757 & 0.2425 & 0.1899 \\
39 & 0.1555 & 0.3259 & 0.2692 & 0.2366 & 0.1860 \\
40 & 0.1503 & 0.3181 & 0.2642 & 0.2325 & 0.1819 \\
41 & 0.1459 & 0.3104 & 0.2585 & 0.2277 & 0.1786 \\
42 & 0.1402 & 0.3015 & 0.2526 & 0.2231 & 0.1758 \\
43 & 0.1356 & 0.2938 & 0.2482 & 0.2192 & 0.1734 \\
44 & 0.1320 & 0.2868 & 0.2439 & 0.2154 & 0.1697 \\
45 & 0.1269 & 0.2792 & 0.2388 & 0.2119 & 0.1656 \\
46 & 0.1223 & 0.2726 & 0.2351 & 0.2082 & 0.1639 \\
47 & 0.1193 & 0.2656 & 0.2306 & 0.2051 & 0.1632 \\
48 & 0.1154 & 0.2601 & 0.2275 & 0.2022 & 0.1600 \\
49 & 0.1116 & 0.2524 & 0.2240 & 0.1989 & 0.1585 \\
50 & 0.1090 & 0.2460 & 0.2194 & 0.1959 & 0.1564 \\
\hline
\end{tabular}
\caption{Numerical results from the left plot of Fig.~\ref{reductionmetricsimage}~(b) in the main text.}
\label{f1scorelowfield}
\end{table}

\begin{table}[h]
\renewcommand{\arraystretch}{1.2}
\begin{tabular}{ |w{c}{2cm}|w{c}{1.8cm}|w{c}{1.8cm}|w{c}{1.8cm}|w{c}{1.8cm}|w{c}{1.8cm}| }
\cline{2-6}
\multicolumn{1}{c|}{} & \multicolumn{5}{c|}{\textbf{MAE of $A^z$ (Hz): low-field regime}} \\ 
\hline
\textbf{Nuclear spins} & \textbf{8 hours} & \textbf{8 hours} & \textbf{6.4 hours} & \textbf{4.8 hours} & \textbf{3.2 hours} \\
\hline
1 & 260.2282 & 405.1872 & 396.5828 & 390.8325 & 384.3432 \\
2 & 251.1654 & 182.7176 & 185.0691 & 187.3382 & 190.0951 \\
3 & 263.6726 & 144.0318 & 148.6022 & 149.6183 & 154.8042 \\
4 & 293.0560 & 147.2674 & 155.4596 & 157.1728 & 160.7815 \\
5 & 329.6712 & 161.3518 & 169.5946 & 173.4622 & 174.8430 \\
6 & 368.8992 & 176.8328 & 190.5612 & 190.6899 & 193.2514 \\
7 & 404.1857 & 191.8234 & 205.8041 & 205.4915 & 209.0273 \\
8 & 443.4137 & 208.1071 & 226.6060 & 227.1567 & 226.2190 \\
9 & 478.4960 & 219.8994 & 245.3275 & 242.6228 & 245.2669 \\
10 & 518.7376 & 232.6298 & 263.0389 & 261.2364 & 264.8836 \\
11 & 549.4742 & 246.0282 & 281.8052 & 280.9841 & 285.0225 \\
12 & 581.7095 & 258.6767 & 298.6365 & 296.5463 & 303.6625 \\
13 & 608.0917 & 271.9054 & 316.6907 & 318.8004 & 328.7630 \\
14 & 631.0861 & 281.2283 & 333.3029 & 335.7496 & 347.7719 \\
15 & 648.6327 & 296.6566 & 352.3886 & 357.3402 & 368.8635 \\
16 & 673.4216 & 308.8433 & 373.2958 & 381.8805 & 396.7316 \\
17 & 681.8454 & 320.9221 & 388.0544 & 399.0695 & 418.3625 \\
18 & 700.6658 & 335.8895 & 411.7712 & 422.2548 & 442.3441 \\
19 & 712.3122 & 347.8367 & 427.5565 & 440.6676 & 468.1088 \\
20 & 721.7626 & 360.6582 & 450.9222 & 462.0290 & 489.9581 \\
21 & 734.6053 & 375.8439 & 467.2184 & 483.7496 & 516.0489 \\
22 & 735.8680 & 385.7603 & 489.0927 & 507.5904 & 539.5339 \\
23 & 747.8732 & 401.8079 & 511.5128 & 531.4620 & 567.1509 \\
24 & 756.6541 & 415.0595 & 530.3218 & 557.9933 & 597.4008 \\
25 & 760.4934 & 423.2896 & 551.6345 & 574.0417 & 620.5144 \\
26 & 768.5827 & 438.0793 & 568.3888 & 594.6057 & 637.5392 \\
27 & 772.6880 & 450.0879 & 589.9317 & 618.7290 & 671.4230 \\
28 & 780.4636 & 462.7080 & 607.4707 & 639.1926 & 693.0527 \\
29 & 788.0663 & 472.2636 & 626.9795 & 656.2526 & 712.8001 \\
30 & 791.1951 & 487.3765 & 649.8318 & 683.6196 & 734.2033 \\
31 & 797.9576 & 494.4483 & 667.8596 & 702.6072 & 759.4413 \\
32 & 803.4013 & 509.5233 & 682.4961 & 724.4530 & 783.1126 \\
33 & 807.7159 & 520.2050 & 708.1764 & 741.4894 & 808.9540 \\
34 & 816.3036 & 531.5855 & 724.5666 & 764.0583 & 826.3106 \\
35 & 826.1163 & 543.7899 & 742.5083 & 782.4291 & 849.5395 \\
36 & 834.3168 & 554.2415 & 760.9621 & 798.1111 & 872.9266 \\
37 & 837.6385 & 565.7272 & 775.1228 & 819.9009 & 887.4358 \\
38 & 844.5581 & 574.8729 & 792.9153 & 839.5465 & 909.1770 \\
39 & 853.1921 & 587.2261 & 808.1229 & 853.7594 & 933.6187 \\
40 & 852.5966 & 597.9441 & 827.3008 & 868.9729 & 946.3203 \\
41 & 860.9590 & 603.1139 & 838.1088 & 888.7731 & 964.0512 \\
42 & 878.5686 & 617.8083 & 854.4698 & 909.1423 & 988.8189 \\
43 & 878.9279 & 629.9923 & 877.2017 & 930.5638 & 1010.2498 \\
44 & 885.5083 & 642.6727 & 899.0703 & 940.4580 & 1022.6586 \\
45 & 894.0113 & 648.4728 & 909.9989 & 953.9354 & 1035.7232 \\
46 & 895.7027 & 658.8026 & 921.4532 & 970.7240 & 1047.3836 \\
47 & 923.8292 & 665.8942 & 934.6619 & 985.8349 & 1072.5958 \\
48 & 923.4695 & 679.2984 & 948.7267 & 1000.3493 & 1077.2146 \\
49 & 923.8201 & 690.8374 & 962.5757 & 1013.0872 & 1086.9868 \\
50 & 939.2969 & 696.6918 & 977.2833 & 1028.6738 & 1108.2456 \\
\hline
\end{tabular}
\caption{Numerical results from the middle plot of Fig.~\ref{reductionmetricsimage}~(b) in the main text.}
\label{maeAlowfield}
\end{table}

\begin{table}[h]
\renewcommand{\arraystretch}{1.2}
\begin{tabular}{ |w{c}{2cm}|w{c}{1.8cm}|w{c}{1.8cm}|w{c}{1.8cm}|w{c}{1.8cm}|w{c}{1.8cm}| }
\cline{2-6}
\multicolumn{1}{c|}{} & \multicolumn{5}{c|}{\textbf{MAE of $A^{\bot}$ (Hz): low-field regime}} \\ 
\hline
\textbf{Nuclear spins} & \textbf{8 hours} & \textbf{8 hours} & \textbf{6.4 hours} & \textbf{4.8 hours} & \textbf{3.2 hours} \\
\hline
1 & 273.8330 & 380.2625 & 374.4611 & 367.6944 & 358.3669 \\
2 & 283.0027 & 235.4731 & 236.8377 & 236.5080 & 238.0131 \\
3 & 297.1525 & 201.2183 & 205.7130 & 205.7466 & 207.8186 \\
4 & 317.3713 & 208.5142 & 214.7527 & 213.3380 & 213.3987 \\
5 & 340.1957 & 225.4855 & 231.0126 & 229.8937 & 226.5270 \\
6 & 367.2399 & 243.2514 & 253.1049 & 247.9331 & 245.6396 \\
7 & 389.2218 & 256.1073 & 267.2818 & 260.8808 & 259.0352 \\
8 & 413.0394 & 274.0824 & 285.5685 & 280.6923 & 271.6389 \\
9 & 431.5511 & 285.2878 & 301.4492 & 293.2219 & 284.8038 \\
10 & 458.5880 & 297.9842 & 316.8277 & 306.3010 & 299.6389 \\
11 & 472.5385 & 307.4897 & 330.9316 & 323.1101 & 311.7457 \\
12 & 493.3198 & 320.9309 & 347.0050 & 336.4758 & 327.9559 \\
13 & 509.1891 & 331.1269 & 363.3086 & 354.8031 & 346.6547 \\
14 & 521.8324 & 341.0750 & 376.3447 & 368.1503 & 363.3562 \\
15 & 539.5297 & 354.5445 & 395.5978 & 383.1231 & 378.9422 \\
16 & 553.7653 & 362.5189 & 409.0987 & 401.6849 & 401.8835 \\
17 & 564.1639 & 377.3383 & 427.4721 & 418.7478 & 422.9385 \\
18 & 576.9198 & 390.4811 & 446.8978 & 439.4152 & 443.9845 \\
19 & 591.7262 & 402.7148 & 461.4408 & 458.7058 & 463.7754 \\
20 & 595.8889 & 413.8770 & 480.8585 & 475.3068 & 485.9680 \\
21 & 602.4835 & 425.4929 & 497.3971 & 500.0390 & 511.4357 \\
22 & 608.1923 & 437.1044 & 518.4661 & 519.2059 & 533.3241 \\
23 & 620.3685 & 452.5285 & 537.8783 & 540.7091 & 563.0763 \\
24 & 627.9462 & 466.4888 & 559.1950 & 562.6841 & 586.2156 \\
25 & 635.0877 & 478.1557 & 577.8161 & 581.9690 & 608.6732 \\
26 & 641.3634 & 492.4579 & 595.1573 & 604.5061 & 631.3480 \\
27 & 649.7685 & 501.8825 & 615.7656 & 625.8019 & 659.7096 \\
28 & 651.4775 & 517.7676 & 631.5600 & 646.4787 & 678.2986 \\
29 & 657.9097 & 526.4128 & 649.9413 & 665.9168 & 699.0617 \\
30 & 658.8850 & 541.2504 & 672.7703 & 691.5914 & 718.9885 \\
31 & 666.2096 & 548.4030 & 691.0638 & 708.6297 & 749.2924 \\
32 & 675.9387 & 567.0924 & 711.4765 & 733.4782 & 772.8354 \\
33 & 679.1565 & 579.0807 & 732.7810 & 751.7591 & 797.2471 \\
34 & 687.1541 & 586.9888 & 747.5967 & 773.6075 & 817.5707 \\
35 & 698.4539 & 600.3888 & 766.6566 & 789.4697 & 839.3250 \\
36 & 692.8538 & 612.8991 & 784.0538 & 814.5438 & 859.2418 \\
37 & 703.2556 & 625.6606 & 804.2034 & 833.0321 & 882.2513 \\
38 & 710.1276 & 632.8623 & 816.4343 & 850.9183 & 904.7801 \\
39 & 719.2829 & 648.0986 & 834.2477 & 865.5570 & 921.9213 \\
40 & 714.8808 & 659.0004 & 851.8566 & 885.3300 & 941.8604 \\
41 & 724.5876 & 669.1622 & 862.1095 & 903.5308 & 956.5774 \\
42 & 740.2988 & 682.0320 & 882.7659 & 921.4523 & 980.2178 \\
43 & 733.6596 & 694.7260 & 900.8817 & 935.6983 & 993.5144 \\
44 & 743.5898 & 708.9141 & 923.3632 & 955.7609 & 1016.3658 \\
45 & 746.4862 & 714.5421 & 934.6434 & 969.7010 & 1025.3020 \\
46 & 744.6608 & 727.1051 & 949.6115 & 981.2866 & 1037.2283 \\
47 & 765.6124 & 734.6673 & 960.2760 & 1000.1270 & 1061.9457 \\
48 & 773.0063 & 746.6041 & 973.8950 & 1014.6178 & 1069.2484 \\
49 & 768.5114 & 757.4871 & 985.8259 & 1027.1274 & 1078.7954 \\
50 & 785.4864 & 765.5762 & 1000.7646 & 1037.4989 & 1096.2875 \\
\hline
\end{tabular}
\caption{Numerical results from the right plot of Fig.~\ref{reductionmetricsimage}~(b) in the main text.}
\label{maeBlowfield}
\end{table}


\begin{thebibliography}{99} 

\bibitem{degen17}
C. L. Degen, F. Reinhard, and P. Cappellaro, Quantum sensing, Rev. Mod. Phys. \textbf{89}, 035002 (2017).

\bibitem{robledo11}
L. Robledo, L. Childress, H. Bernien, B. Hensen, P. F. Alkemade, and R. Hanson, High-fidelity projective read-out of a solid-state spin quantum register, Nat. \textbf{477}, 574-578 (2011).

\bibitem{zhang21}
Q. Zhang, Y. Guo, W. Ji, M. Wang, J. Yin, F. Kong, Y. Lin, C. Yin, F. Shi, Y. Wang, and J. Du, High-fidelity single-shot readout of single electron spin in diamond with spin-to-charge conversion, Nat. Commun. \textbf{12}, 1529 (2021).

\bibitem{awschalom18}
D. D. Awschalom, R. Hanson, J. Wrachtrup, and B. B. Zhou, Quantum technologies with optically interfaced solid-state spins, Nat. Photonics \textbf{12}, 516-527 (2018).

\bibitem{gebhart23}
V. Gebhart, R. Santagati, A. A. Gentile, E. M. Gauger, D. Craig, N. Ares, L Banchi, F. Marquardt, L. Pezzè, and C. Bonato, Learning quantum systems, Nat. Rev. Phys. \textbf{5}, 141-156 (2023).

\bibitem{childress13}
L. Childress, and R. Hanson, Diamond NV centers for quantum computing and quantum networks, MRS Bull. \textbf{38}, 134-138 (2013).

\bibitem{tisler13}
J. Tisler, T. Oeckinghaus, R. J. Stohr, R. Kolesov, R. Reuter, F. Reinhard, and J. Wrachtrup, Single Defect Center Scanning Near-Field Optical Microscopy on Graphene, Nano Lett. \textbf{13}, 3152-3156 (2013).

\bibitem{nielsen17}
M. A. Nielsen, and I. L. Chuang, Quantum Computation and Quantum Information, Cambridge University Press (2010).

\bibitem{vorobyov21}
V. Vorobyov, J. Javadzade, M. Joliffe, F. Kaiser, and J. Wrachtrup, Addressing Single Nuclear Spins Quantum Memories by a Central Electron Spin, Appl. Magn. Reson. \textbf{53}, 1317-1330 (2022).

\bibitem{budakian24}
R. Budakian, A. Finkler, A. Eichler, M. Poggio, C. L. Degen, s. Tabatabaei, 
I. Lee, P. C. Hammel, S. P. Eugene, T. H. Taminiau, R. L. Walsworth, P. London, A. B. Jayich, A. Ajoy, A. Pillai, J. Wrachtrup, F. Jelezko, Y. Bae, A. J. Heinrich, C. R. Ast, P. Bertet, P. Cappellaro, C. Bonato, Y. Altmann, and E. Gauger, Roadmap on nanoscale magnetic resonance imaging, Nanotechnol. \textbf{35}, 412001 (2024).

\bibitem{du24}
J. Du, F. Shi, X. Kong, F. Jelezko, and J. Wrachtrup, Single-molecule scale magnetic resonance spectroscopy using quantum diamond sensors, Rev. Mod. Phys. \textbf{96}, 025001 (2024).

\bibitem{schwartz19}
I. Schwartz, J. Rosskopf, S. Schmitt, B. Tratzmiller, Q. Chen, L. P. McGuinness, F. Jelezko, and M. B. Plenio, Blueprint for nanoscale NMR, Sci. Rep. \textbf{9}, 6938 (2019).

\bibitem{cramer16}
J. Cramer, N. Kalb, M. A. Rol, B. Hensen, M. S. Blok, M. Markham, D. J. Twitchen, R. Hanson, and T. H. Taminiau, Repeated quantum error correction on a continuously encoded qubit by real-time feedback, Nat. Commun. \textbf{7}, 11526 (2016).

\bibitem{abobeih22} 
M. H. Abobeih, Y. Wang, J. Randall, S. J. H. Loenen, C. E. Bradley, M. Markham, D. J. Twitchen, B. M. Terhal, and T. H. Taminiau Fault-tolerant operation of a logical qubit in a diamond quantum processor, Nat. \textbf{606}, 884-889 (2022).

\bibitem{reiner24}
J. Reiner, Y. Chung, S. H. Misha, C. Lehner, C. Moehle, D. Poulos, S. Monir, K. J. Charde, P. Macha, L. Kranz, I. Thorvaldson, B. Thorgrimsson, D. Keith, Y. L. Hsueh, R. Rahman, S. K. Gorman, J. G. Keizer, and M. Y. Simmons, High-fidelity initialization and control of electron and nuclear spins in a four-qubit register, Nat. Nanotechnol. \textbf{19}, 605-611 (2024).

\bibitem{bradley22}
C. E. Bradley, S. W. de Bone, P. F. W. Möller, S. Baier, M. J. Degen, S. J. H. Loenen, H. P. Bartling, M.Markham, D. J. Twitchen, R. Hanson, D. Elkouss, and T. H. Taminiau, Robust quantum-network memory based on spin qubits in isotopically engineered diamond, Npj Quantum Inf. \textbf{8}, 122 (2022).

\bibitem{vanommen23}
H. B. van Ommen, G. L. van de Stolpe, N. Demetriou, H. K. C. Beukers, J. Yun, T. R. J. Fortuin, M. Iuliano, A. R.-P. Montblanch, R. Hanson, and T. H. Taminiau, Improved Electron-Nuclear Quantum Gates for Spin Sensing and Control, PRX Quantum \textbf{6}, 020309 (2025).

\bibitem{maity22}
S. Maity, B. Pingault, G. Joe, M. Chalupnik, D. Assumpção, E. Cornell, L. Shao, and M. Lončar, Mechanical control of a single nuclear spin, Phys. Rev. X \textbf{12}, 011056 (2022).

\bibitem{beukers25}
H. K. Beukers, C. Waas, M. Pasini, H. B. Van Ommen, Z. Ademi, M. Iuliano, N. Codreanu, J. M. Brevoord, T. Turan, T. H. Taminiau, and R. Hanson, Control of solid-state nuclear spin qubits using an electron spin-1/2, Phys. Rev. X \textbf{15}, 021011 (2025).

\bibitem{bourassa20}
A. Bourassa, C. P. Anderson, K. C. Miao, M. Onizhuk, H. Ma, A. L. Crook, H. Abe, J. Ul-Hassan, T. Ohshima, N. T. Son, G. Galli, and D. D. Awschalom, Entanglement and control of single nuclear spins in isotopically engineered silicon carbide, Nat. Mater. \textbf{19}, 1319-1325 (2020).

\bibitem{zhao12}
S. Kolkowitz, Q. P. Unterreithmeier, S. D. Bennett, and M. D. Lukin,  Sensing Distant Nuclear Spins with a Single Electron Spin, Phys. Rev. Lett. \textbf{109}, 137601 (2012).

\bibitem{kolkowitz12}
N. Zhao, J. Honert, B. Schmid, M. Klas, J. Isoya, M. Markham, D. Twitchen, F. Jelezko, R. B. Liu, H. Fedder, and J. Wrachtrup, Sensing single remote nuclear spins, Nat. Nanotechnol. \textbf{7}, 657-662 (2012).

\bibitem{taminiau12}
T. H. Taminiau, J. J. T. Wagenaar, T. Van der Sar, F. Jelezko, V. V. Dobrovitski, and R. Hanson, Detection and Control of Individual Nuclear Spins using a Weakly Coupled Electron Spin, Phys. Rev. Lett. \textbf{109}, 137602 (2012).

\bibitem{fazhan14}
F. Shi, X. Kong, P. Wang, F. Kong, N. Zhao, R. B. Liu, and J. Du, Sensing and atomic-scale structure analysis of single nuclear-spin clusters in diamond, Nat. Phys. \textbf{10}, 21-25 (2014).

\bibitem{muller14}
C. Müller, X. Kong, J. M. Cai, K. Melentijević, A. Stacey, M. Markham, D. Twitchen, J. Isoya, S. Pezzagna, J. Meijer, J. F. Du, M. B. Plenio, B. Naydenov, L. P. McGuinness, and F. Jelezko, Nuclear magnetic resonance spectroscopy with single spin sensitivity, Nat. Commun. \textbf{5}, 4703 (2014).

\bibitem{liu16}
G. Q. Liu, J. Xing, W. L. Ma, P. Wang, C. H. Li, H. C. Po, Y. R. Zhang, H. Fan, R. B. Liu, and X. Y. Pan, Single-Shot Readout of a Nuclear Spin Weakly Coupled to a Nitrogen-Vacancy Center at Room Temperature, Phys. Rev. Lett. \textbf{118}, 150504 (2017).

\bibitem{zopes18}
J. Zopes, K. S. Cujia, K. Sasaki, J. M. Boss, K. M. Itoh, and C. L. Degen, Three-dimensional localization spectroscopy of individual nuclear spins with sub-Angstrom resolution, Nat. Commun. \textbf{9}, 4678 (2018).

\bibitem{abobeih18}
M. H. Abobeih, J. Cramer, M. A. Bakker, N. Kalb, M. Markham, D. J. Twitchen, and T. H. Taminiau, One-second coherence for a single electron spin coupled to a multi-qubit nuclear-spin environment, Nat. Commun. \textbf{9}, 2552 (2018).

\bibitem{abobeih19}
M. H. Abobeih, J. Randall, C. E. Bradley, H. P. Bartling, M. A. Bakker, M. J. Degen, M. Markham, D. J. Twitchen, and T. H. Taminiau, Atomic-scale imaging of a 27-nuclear-spin cluster using a quantum sensor, Nature \textbf{576}, 411-415 (2019).

\bibitem{jung21}
K. Jung, M. H. Abobeih, J. Yun, G. Kim, H. Oh, A. Henry, T. H. Taminiau, and D. Kim, Deep learning enhanced individual nuclear-spin detection, Npj Quantum Inf. \textbf{7}, 41 (2021).

\bibitem{galli25}
A. N. Poteshman, M. Onizhuk, C. Egerstrom, D. P. Mark, D. D. Awschalom, F. J. Heremans, and G. Galli, High-throughput spin-bath characterization of spin-defects in semiconductors, Phys. Rev. Appl. \textbf{24}, 054048 (2025).

\bibitem{bradley19}
C. E. Bradley, J. Randall, M. H. Abobeih, R. C. Berrevoets, M. J. Degen, M. A. Bakker, M. Markham, D. J. Twitchen, and T. H. Taminiau, A Ten-Qubit Solid-State Spin Register with Quantum Memory up to One Minute, Phys. Rev. X \textbf{9}, 031045 (2019).

\bibitem{zahedian24}
M. Zahedian, V. Vorobyov, and J. Wrachtrup, Blueprint for efficient nuclear spin characterization with color centers, Phys. Rev. B \textbf{109}, 214111 (2024).


\bibitem{joas21}
T. Joas, S. Schmitt, R. Santagati, A. A. Gentile, C. Bonato, A. Laing, L. P. McGuinness, and F. Jelezko, Online adaptive quantum characterization of a nuclear spin, Npj Quantum Inf. \textbf{7}, 56 (2021).

\bibitem{arshad24}
M. J. Arshad, C. Bekker, B. Haylock, K. Skrzypczak, D. White, B. Griffiths, J. Gore, G. W. Morley, P. Salter, J. Smith, I. Zohar, A. Finkler, Y. Altmann, E. M. Gauger, and C. Bonato, Real-time adaptive estimation of decoherence timescales for a single qubit, Phys. Rev. Appl. \textbf{21}, 024026 (2024).

\bibitem{belliardo23} F. Belliardo, F. Zoratti, F. Marquardt, and V. Giovannetti, Model-aware reinforcement learning for high-performance Bayesian experimental design in quantum metrology, Quantum \textbf{8}, 1555 (2024).

\bibitem{belliardo24} F. Belliardo, F. Zoratti, and V. Giovannetti, Applications of model-aware reinforcement learning in Bayesian quantum metrology, Phys. Rev. A \textbf{109}, 062609 (2024).

\bibitem{varona24}
B. Varona-Uriarte, C. Munuera-Javaloy, E. Terradillos, Y. Ban, A. Alvarez-Gila, E. Garrote, and J. Casanova, Automatic Detection of Nuclear Spins at Arbitrary Magnetic Fields via Signal-to-Image AI Model, Phys. Rev. Lett. {\bf 132}, 150801 (2024).


\bibitem{belliardo25}
F. Belliardo, E. M. Gauger, M. H. Abobeih, T. H. Taminiau, Y. Altmann, and C. Bonato, Multidimensional quantum estimation and model learning framework based on variational Bayesian inference, accepted in PRX Quantum (2026).


\bibitem{shi14}
F. Shi, X. Kong, P. Wang, F. Kong, N. Zhao, R. B. Liu, and J. Du, Sensing and atomic-scale structure analysis of single nuclear-spin clusters in diamond, Nat. Phys. \textbf{10}, 21-25 (2014).


\bibitem{lindley56}
D. V. Lindley, On a measure of the information provided by an experiment, Ann. Math. Statist. \textbf{27}, 986 (1956).


\bibitem{rainforth24}
T. Rainforth, A. Foster, D. R. Ivanova, and F. Bickford Smith, Modern Bayesian experimental design, Statis. Sci. \textbf{39}, 100 (2024).


\bibitem{sarra23}
L. Sarra, and F. Marquardt, Deep Bayesian experimental design for quantum many-body systems, Mach. Learn. Sci. Technol.
\textbf{4}, 045022 (2023).


\bibitem{vincent17}
B. T. Vincent, and T. Rainforth, The DARC Toolbox: automated, flexible, and efficient delayed and risky choice experiments using Bayesian adaptive design, PsyArXiv:10.31234 (2017).


\bibitem{foster21}
A. E. Foster, Variational, Monte Carlo and policy-based approaches to Bayesian experimental design, Ph.D. thesis, University of Oxford (2021).


\bibitem{supplementalmaterial}
Supplemental Material for Letter \textit{Computationally Tractable Offline Quantum Experimental Design for Nuclear Spin Detection}.

\bibitem{jing23}
L. Jing, P. Du, H. Tang, and W. Zhang, Noise-resistant quantum memory enabled by Hamiltonian engineering, Phys. Rev. A \textbf{107}, 012601 (2023).

\bibitem{dobrovitski06}
V. V. Dobrovitski, J. M. Taylor, and M. D. Lukin, Long-lived memory for electronic spin in a quantum dot: Numerical analysis, Phys. Rev. B \textbf{73}, 245318 (2006).

\bibitem{waeber19}
A. M. Waeber, G. Gillard, G. Ragunathan, M. Hopkinson, P. Spencer, D. A. Ritchie, M. S.Skolnick, and E. A. Chekhovich, Pulse control protocols for preserving coherence in dipolar-coupled nuclear spin baths, Nat. Commun. \textbf{10}, 3157 (2019).

\bibitem{marcks25}
J. C. Marcks, B. Pingault, J. Zhang, C. Zeledon, F. J. Heremans, and D. D. Awschalom, Nuclear spin engineering for quantum information science, J. Mater. Res., 1-16 (2025).

\bibitem{briegel25}
K. D. Briegel, N. R. von Grafenstein, J. C. Draeger, P. Blümler, R. D. Allert, and D. B. Bucher, Optical widefield nuclear magnetic resonance microscopy, Nat. Commun. \textbf{16}, 1281 (2025).

\bibitem{aslam17}
N. Aslam, M. Pfender, P. Neumann, R. Reuter, A. Zappe, F. Fávaro de Oliveira, A. Denisenko, H. Sumiya, S. Onoda, J. Isoya, and J. Wrachtrup, Nanoscale nuclear magnetic resonance with chemical resolution, Science \textbf{357}, 67-71 (2017).

\bibitem{aslam23}
N. Aslam, H. Zhou, E. K. Urbach, M. J. Turner, R. L. Walsworth, M. D. Lukin, and H. Park, Quantum sensors for biomedical applications, Nat. Rev. Phys. \textbf{5}, 157-169 (2023).


\bibitem{valeri23}
M. Valeri, V. Cimini, S. Piacentini, F. Ceccarelli, E. Polino, F. Hoch, G. Bizzarri, G. Corrielli, N. Spagnolo, R. Osellame, and F. Sciarrino, Experimental multiparameter quantum metrology in adaptive regime, Phys. Rev. Res. \textbf{5}, 013138 (2023).

\bibitem{valeri20}
M. Valeri, E. Polino, D. Poderini, S. Giacomini, G. Corrielli, A. Crespi, R. Osellame, N. Spagnolo, and F. Sciarrino, Experimental adaptive Bayesian estimation of multiple phases with limited data, npj Quantum Inf. \textbf{6}, 92 (2020).

\bibitem{tsang11}
M. Tsang, H. M. Wiseman, and C. M. Caves, Fundamental quantum limit to waveform estimation, Phys. Rev. Lett. \textbf{106}, 090401 (2011).

\bibitem{gardner25}
J. W. Gardner, T. Gefen, S. A. Haine, J. J. Hope, J. Preskill, Y. Chen, and L. McCuller, Stochastic waveform estimation at the fundamental quantum limit, PRX Quantum \textbf{6}, 030311 (2025).


\end{thebibliography}

\begin{thebibliography}{99} 

\bibitem{staminiau12}
T. H. Taminiau, J. J. T. Wagenaar, T. Van der Sar, F. Jelezko, V. V. Dobrovitski, and R. Hanson, Detection and control of individual nuclear spins using a weakly coupled electron spin, Phys. Rev. Lett. \textbf{109}, 137602 (2012).

\bibitem{svarona24}
B. Varona-Uriarte, C. Munuera-Javaloy, E. Terradillos, Y. Ban, A. Alvarez-Gila, E. Garrote, and J. Casanova, Automatic Detection of Nuclear Spins at Arbitrary Magnetic Fields via Signal-to-Image AI Model, Phys. Rev. Lett. {\bf 132}, 150801 (2024).

\bibitem{sjung21}
K. Jung, M. H. Abobeih, J. Yun, G. Kim, H. Oh, A. Henry, T. H. Taminiau, and D. Kim, Deep learning enhanced individual nuclear-spin detection, Npj Quantum Inf. \textbf{7}, 41 (2021).



\end{thebibliography}
\end{document}